# The Pierre Auger Observatory II: Studies of Cosmic Ray Composition and Hadronic Interaction models

THE PIERRE AUGER COLLABORATION

*Observatorio Pierre Auger, Av. San Martín Norte 304, 5613 Malargüe, Argentina*

# The Pierre Auger Collaboration

P. Abreu[74], M. Aglietta[57], E.J. Ahn[93], I.F.M. Albuquerque[19], D. Allard[33], I. Allekotte[1], J. Allen[96], P. Allison[98], J. Alvarez Castillo[67], J. Alvarez-Muñiz[84], M. Ambrosio[50], A. Aminaei[68], L. Anchordoqui[109], S. Andringa[74], T. Antičić[27], A. Anzalone[56], C. Aramo[50], E. Arganda[81], F. Arqueros[81], H. Asorey[1], P. Assis[74], J. Aublin[35], M. Ave[41], M. Avenier[36], G. Avila[12], T. Bäcker[45], M. Balzer[40], K.B. Barber[13], A.F. Barbosa[16], R. Bardenet[34], S.L.C. Barroso[22], B. Baughman[98], J. Bäuml[39], J.J. Beatty[98], B.R. Becker[106], K.H. Becker[38], A. Bellétoile[37], J.A. Bellido[13], S. BenZvi[108], C. Berat[36], X. Bertou[1], P.L. Biermann[42], P. Billoir[35], F. Blanco[81], M. Blanco[82], C. Bleve[38], H. Blümer[41, 39], M. Boháčová[29, 101], D. Boncioli[51], C. Bonifazi[25, 35], R. Bonino[57], N. Borodai[72], J. Brack[91], P. Brogueira[74], W.C. Brown[92], R. Bruijn[87], P. Buchholz[45], A. Bueno[83], R.E. Burton[89], K.S. Caballero-Mora[99], L. Caramete[42], R. Caruso[52], A. Castellina[57], O. Catalano[56], G. Cataldi[49], L. Cazon[74], R. Cester[53], J. Chauvin[36], S.H. Cheng[99], A. Chiavassa[57], J.A. Chinellato[20], A. Chou[93, 96], J. Chudoba[29], R.W. Clay[13], M.R. Coluccia[49], R. Conceição[74], F. Contreras[11], H. Cook[87], M.J. Cooper[13], J. Coppens[68, 70], A. Cordier[34], U. Cotti[66], S. Coutu[99], C.E. Covault[89], A. Creusot[33, 79], A. Criss[99], J. Cronin[101], A. Curutiu[42], S. Dagoret-Campagne[34], R. Dallier[37], S. Dasso[8, 4], K. Daumiller[39], B.R. Dawson[13], R.M. de Almeida[26], M. De Domenico[52], C. De Donato[67, 48], S.J. de Jong[68, 70], G. De La Vega[10], W.J.M. de Mello Junior[20], J.R.T. de Mello Neto[25], I. De Mitri[49], V. de Souza[18], K.D. de Vries[69], G. Decerprit[33], L. del Peral[82], O. Deligny[32], H. Dembinski[41], N. Dhital[95], C. Di Giulio[47, 51], J.C. Diaz[95], M.L. Díaz Castro[17], P.N. Diep[110], C. Dobrigkeit [20], W. Docters[69], J.C. D'Olivo[67], P.N. Dong[110, 32], A. Dorofeev[91], J.C. dos Anjos[16], M.T. Dova[7], D. D'Urso[50], I. Dutan[42], J. Ebr[29], R. Engel[39], M. Erdmann[43], C.O. Escobar[20], A. Etchegoyen[2], P. Facal San Luis[101], I. Fajardo Tapia[67], H. Falcke[68, 71], G. Farrar[96], A.C. Fauth[20], N. Fazzini[93], A.P. Ferguson[89], A. Ferrero[2], B. Fick[95], A. Filevich[2], A. Filipčič[78, 79], S. Fliescher[43], C.E. Fracchiolla[91], E.D. Fraenkel[69], U. Fröhlich[45], B. Fuchs[16], R. Gaior[35], R.F. Gamarra[2], S. Gambetta[46], B. García[10], D. García Gámez[83], D. Garcia-Pinto[81], A. Gascon[83], H. Gemmeke[40], K. Gesterling[106], P.L. Ghia[35, 57], U. Giaccari[49], M. Giller[73], H. Glass[93], M.S. Gold[106], G. Golup[1], F. Gomez Albarracin[7], M. Gómez Berisso[1], P. Gonçalves[74], D. Gonzalez[41], J.G. Gonzalez[41], B. Gookin[91], D. Góra[41, 72], A. Gorgi[57], P. Gouffon[19], S.R. Gozzini[87], E. Grashorn[98], S. Grebe[68, 70], N. Griffith[98], M. Grigat[43], A.F. Grillo[58], Y. Guardincerri[4], F. Guarino[50], G.P. Guedes[21], A. Guzman[67], J.D. Hague[106], P. Hansen[7], D. Harari[1], S. Harmsma[69, 70], J.L. Harton[91], A. Haungs[39], T. Hebbeker[43], D. Heck[39], A.E. Herve[13], C. Hojvat[93], N. Hollon[101], V.C. Holmes[13], P. Homola[72], J.R. Hörandel[68], A. Horneffer[68], M. Hrabovský[30, 29], T. Huege[39], A. Insolia[52], F. Ionita[101], A. Italiano[52], C. Jarne[7], S. Jiraskova[68], M. Josebachuili[2], K. Kadija[27], K.-H. Kampert[38], P. Karhan[28], P. Kasper[93], B. Kégl[34], B. Keilhauer[39], A. Keivani[94], J.L. Kelley[68], E. Kemp[20], R.M. Kieckhafer[95], H.O. Klages[39], M. Kleifges[40], J. Kleinfeller[39], J. Knapp[87], D.-H. Koang[36], K. Kotera[101], N. Krohm[38], O. Krömer[40], D. Kruppke-Hansen[38], F. Kuehn[93], D. Kuempel[38], J.K. Kulbartz[44], N. Kunka[40], G. La Rosa[56], C. Lachaud[33], P. Lautridou[37], M.S.A.B. Leão[24], D. Lebrun[36], P. Lebrun[93], M.A. Leigui de Oliveira[24], A. Lemiere[32], A. Letessier-Selvon[35], I. Lhenry-Yvon[32], K. Link[41], R. López[63], A. Lopez Agüera[84], K. Louedec[34], J. Lozano Bahilo[83], A. Lucero[2, 57], M. Ludwig[41], H. Lyberis[32], M.C. Maccarone[56], C. Macolino[35], S. Maldera[57], D. Mandat[29], P. Mantsch[93], A.G. Mariazzi[7], J. Marin[11, 57], V. Marin[37], I.C. Maris[35], H.R. Marquez Falcon[66], G. Marsella[54], D. Martello[49], L. Martin[37], H. Martinez[64], O. Martínez Bravo[63],

H.J. Mathes[39], J. Matthews[94, 100], J.A.J. Matthews[106], G. Matthiae[51], D. Maurizio[53], P.O. Mazur[93], G. Medina-Tanco[67], M. Melissas[41], D. Melo[2, 53], E. Menichetti[53], A. Menshikov[40], P. Mertsch[85], C. Meurer[43], S. Mićanović[27], M.I. Micheletti[9], W. Miller[106], L. Miramonti[48], S. Mollerach[1], M. Monasor[101], D. Monnier Ragaigne[34], F. Montanet[36], B. Morales[67], C. Morello[57], E. Moreno[63], J.C. Moreno[7], C. Morris[98], M. Mostafá[91], C.A. Moura[24, 50], S. Mueller[39], M.A. Muller[20], G. Müller[43], M. Münchmeyer[35], R. Mussa[53], G. Navarra[57 †], J.L. Navarro[83], S. Navas[83], P. Necesal[29], L. Nellen[67], A. Nelles[68, 70], J. Neuser[38], P.T. Nhung[110], L. Niemietz[38], N. Nierstenhoefer[38], D. Nitz[95], D. Nosek[28], L. Nožka[29], M. Nyklicek[29], J. Oehlschläger[39], A. Olinto[101], V.M. Olmos-Gilbaja[84], M. Ortiz[81], N. Pacheco[82], D. Pakk Selmi-Dei[20], M. Palatka[29], J. Pallotta[3], N. Palmieri[41], G. Parente[84], E. Parizot[33], A. Parra[84], R.D. Parsons[87], S. Pastor[80], T. Paul[97], M. Pech[29], J. Pękala[72], R. Pelayo[84], I.M. Pepe[23], L. Perrone[54], R. Pesce[46], E. Petermann[105], S. Petrera[47], P. Petrinca[51], A. Petrolini[46], Y. Petrov[91], J. Petrovic[70], C. Pfendner[108], N. Phan[106], R. Piegaia[4], T. Pierog[39], P. Pieroni[4], M. Pimenta[74], V. Pirronello[52], M. Platino[2], V.H. Ponce[1], M. Pontz[45], P. Privitera[101], M. Prouza[29], E.J. Quel[3], S. Querchfeld[38], J. Rautenberg[38], O. Ravel[37], D. Ravignani[2], B. Revenu[37], J. Ridky[29], S. Riggi[84, 52], M. Risse[45], P. Ristori[3], H. Rivera[48], V. Rizi[47], J. Roberts[96], C. Robledo[63], W. Rodrigues de Carvalho[84, 19], G. Rodriguez[84], J. Rodriguez Martino[11, 52], J. Rodriguez Rojo[11], I. Rodriguez-Cabo[84], M.D. Rodríguez-Frías[82], G. Ros[82], J. Rosado[81], T. Rossler[30], M. Roth[39], B. Rouillé-d'Orfeuil[101], E. Roulet[1], A.C. Rovero[8], C. Rühle[40], F. Salamida[47, 39], H. Salazar[63], G. Salina[51], F. Sánchez[2], M. Santander[11], C.E. Santo[74], E. Santos[74], E.M. Santos[25], F. Sarazin[90], B. Sarkar[38], S. Sarkar[85], R. Sato[11], N. Scharf[43], V. Scherini[48], H. Schieler[39], P. Schiffer[43], A. Schmidt[40], F. Schmidt[101], O. Scholten[69], H. Schoorlemmer[68, 70], J. Schovancova[29], P. Schovánek[29], F. Schröder[39], S. Schulte[43], D. Schuster[90], S.J. Sciutto[7], M. Scuderi[52], A. Segreto[56], M. Settimo[45], A. Shadkam[94], R.C. Shellard[16, 17], I. Sidelnik[2], G. Sigl[44], H.H. Silva Lopez[67], A. Śmiałkowski[73], R. Šmída[39, 29], G.R. Snow[105], P. Sommers[99], J. Sorokin[13], H. Spinka[88, 93], R. Squartini[11], S. Stanic[79], J. Stapleton[98], J. Stasielak[72], M. Stephan[43], E. Strazzeri[56], A. Stutz[36], F. Suarez[2], T. Suomijärvi[32], A.D. Supanitsky[8, 67], T. Šuša[27], M.S. Sutherland[94, 98], J. Swain[97], Z. Szadkowski[73], M. Szuba[39], A. Tamashiro[8], A. Tapia[2], M. Tartare[36], O. Taşcău[38], C.G. Tavera Ruiz[67], R. Tcaciuc[45], D. Tegolo[52, 61], N.T. Thao[110], D. Thomas[91], J. Tiffenberg[4], C. Timmermans[70, 68], D.K. Tiwari[66], W. Tkaczyk[73], C.J. Todero Peixoto[18, 24], B. Tomé[74], A. Tonachini[53], P. Travnicek[29], D.B. Tridapalli[19], G. Tristram[33], E. Trovato[52], M. Tueros[84, 4], R. Ulrich[99, 39], M. Unger[39], M. Urban[34], J.F. Valdés Galicia[67], I. Valiño[84, 39], L. Valore[50], A.M. van den Berg[69], E. Varela[63], B. Vargas Cárdenas[67], J.R. Vázquez[81], R.A. Vázquez[84], D. Veberič[79, 78], V. Verzi[51], J. Vicha[29], M. Videla[10], L. Villaseñor[66], H. Wahlberg[7], P. Wahrlich[13], O. Wainberg[2], D. Walz[43], D. Warner[91], A.A. Watson[87], M. Weber[40], K. Weidenhaupt[43], A. Weindl[39], S. Westerhoff[108], B.J. Whelan[13], G. Wieczorek[73], L. Wiencke[90], B. Wilczyńska[72], H. Wilczyński[72], M. Will[39], C. Williams[101], T. Winchen[43], L. Winders[109], M.G. Winnick[13], M. Wommer[39], B. Wundheiler[2], T. Yamamoto[101 a], T. Yapici[95], P. Younk[45], G. Yuan[94], A. Yushkov[84, 50], B. Zamorano[83], E. Zas[84], D. Zavrtanik[79, 78], M. Zavrtanik[78, 79], I. Zaw[96], A. Zepeda[64], M. Zimbres-Silva[20, 38] M. Ziolkowski[45]

[1] *Centro Atómico Bariloche and Instituto Balseiro (CNEA- UNCuyo-CONICET), San Carlos de Bariloche, Argentina*
[2] *Centro Atómico Constituyentes (Comisión Nacional de Energía Atómica/CONICET/UTN-FRBA), Buenos Aires, Argentina*
[3] *Centro de Investigaciones en Láseres y Aplicaciones, CITEFA and CONICET, Argentina*
[4] *Departamento de Física, FCEyN, Universidad de Buenos Aires y CONICET, Argentina*
[7] *IFLP, Universidad Nacional de La Plata and CONICET, La Plata, Argentina*
[8] *Instituto de Astronomía y Física del Espacio (CONICET- UBA), Buenos Aires, Argentina*
[9] *Instituto de Física de Rosario (IFIR) - CONICET/U.N.R. and Facultad de Ciencias Bioquímicas y Farmacéuticas U.N.R., Rosario, Argentina*
[10] *National Technological University, Faculty Mendoza (CONICET/CNEA), Mendoza, Argentina*
[11] *Observatorio Pierre Auger, Malargüe, Argentina*
[12] *Observatorio Pierre Auger and Comisión Nacional de Energía Atómica, Malargüe, Argentina*
[13] *University of Adelaide, Adelaide, S.A., Australia*
[16] *Centro Brasileiro de Pesquisas Fisicas, Rio de Janeiro, RJ, Brazil*
[17] *Pontifícia Universidade Católica, Rio de Janeiro, RJ, Brazil*

[18] *Universidade de São Paulo, Instituto de Física, São Carlos, SP, Brazil*
[19] *Universidade de São Paulo, Instituto de Física, São Paulo, SP, Brazil*
[20] *Universidade Estadual de Campinas, IFGW, Campinas, SP, Brazil*
[21] *Universidade Estadual de Feira de Santana, Brazil*
[22] *Universidade Estadual do Sudoeste da Bahia, Vitoria da Conquista, BA, Brazil*
[23] *Universidade Federal da Bahia, Salvador, BA, Brazil*
[24] *Universidade Federal do ABC, Santo André, SP, Brazil*
[25] *Universidade Federal do Rio de Janeiro, Instituto de Física, Rio de Janeiro, RJ, Brazil*
[26] *Universidade Federal Fluminense, EEIMVR, Volta Redonda, RJ, Brazil*
[27] *Rudjer Bošković Institute, 10000 Zagreb, Croatia*
[28] *Charles University, Faculty of Mathematics and Physics, Institute of Particle and Nuclear Physics, Prague, Czech Republic*
[29] *Institute of Physics of the Academy of Sciences of the Czech Republic, Prague, Czech Republic*
[30] *Palacky University, RCATM, Olomouc, Czech Republic*
[32] *Institut de Physique Nucléaire d'Orsay (IPNO), Université Paris 11, CNRS-IN2P3, Orsay, France*
[33] *Laboratoire AstroParticule et Cosmologie (APC), Université Paris 7, CNRS-IN2P3, Paris, France*
[34] *Laboratoire de l'Accélérateur Linéaire (LAL), Université Paris 11, CNRS-IN2P3, Orsay, France*
[35] *Laboratoire de Physique Nucléaire et de Hautes Energies (LPNHE), Universités Paris 6 et Paris 7, CNRS-IN2P3, Paris, France*
[36] *Laboratoire de Physique Subatomique et de Cosmologie (LPSC), Université Joseph Fourier, INPG, CNRS-IN2P3, Grenoble, France*
[37] *SUBATECH, École des Mines de Nantes, CNRS-IN2P3, Université de Nantes, Nantes, France*
[38] *Bergische Universität Wuppertal, Wuppertal, Germany*
[39] *Karlsruhe Institute of Technology - Campus North - Institut für Kernphysik, Karlsruhe, Germany*
[40] *Karlsruhe Institute of Technology - Campus North - Institut für Prozessdatenverarbeitung und Elektronik, Karlsruhe, Germany*
[41] *Karlsruhe Institute of Technology - Campus South - Institut für Experimentelle Kernphysik (IEKP), Karlsruhe, Germany*
[42] *Max-Planck-Institut für Radioastronomie, Bonn, Germany*
[43] *RWTH Aachen University, III. Physikalisches Institut A, Aachen, Germany*
[44] *Universität Hamburg, Hamburg, Germany*
[45] *Universität Siegen, Siegen, Germany*
[46] *Dipartimento di Fisica dell'Università and INFN, Genova, Italy*
[47] *Università dell'Aquila and INFN, L'Aquila, Italy*
[48] *Università di Milano and Sezione INFN, Milan, Italy*
[49] *Dipartimento di Fisica dell'Università del Salento and Sezione INFN, Lecce, Italy*
[50] *Università di Napoli "Federico II" and Sezione INFN, Napoli, Italy*
[51] *Università di Roma II "Tor Vergata" and Sezione INFN, Roma, Italy*
[52] *Università di Catania and Sezione INFN, Catania, Italy*
[53] *Università di Torino and Sezione INFN, Torino, Italy*
[54] *Dipartimento di Ingegneria dell'Innovazione dell'Università del Salento and Sezione INFN, Lecce, Italy*
[56] *Istituto di Astrofisica Spaziale e Fisica Cosmica di Palermo (INAF), Palermo, Italy*
[57] *Istituto di Fisica dello Spazio Interplanetario (INAF), Università di Torino and Sezione INFN, Torino, Italy*
[58] *INFN, Laboratori Nazionali del Gran Sasso, Assergi (L'Aquila), Italy*
[61] *Università di Palermo and Sezione INFN, Catania, Italy*
[63] *Benemérita Universidad Autónoma de Puebla, Puebla, Mexico*
[64] *Centro de Investigación y de Estudios Avanzados del IPN (CINVESTAV), México, D.F., Mexico*
[66] *Universidad Michoacana de San Nicolas de Hidalgo, Morelia, Michoacan, Mexico*
[67] *Universidad Nacional Autonoma de Mexico, Mexico, D.F., Mexico*
[68] *IMAPP, Radboud University Nijmegen, Netherlands*
[69] *Kernfysisch Versneller Instituut, University of Groningen, Groningen, Netherlands*
[70] *Nikhef, Science Park, Amsterdam, Netherlands*
[71] *ASTRON, Dwingeloo, Netherlands*
[72] *Institute of Nuclear Physics PAN, Krakow, Poland*

[73] University of Łódź, Łódź, Poland
[74] LIP and Instituto Superior Técnico, Lisboa, Portugal
[78] J. Stefan Institute, Ljubljana, Slovenia
[79] Laboratory for Astroparticle Physics, University of Nova Gorica, Slovenia
[80] Instituto de Física Corpuscular, CSIC-Universitat de València, Valencia, Spain
[81] Universidad Complutense de Madrid, Madrid, Spain
[82] Universidad de Alcalá, Alcalá de Henares (Madrid), Spain
[83] Universidad de Granada & C.A.F.P.E., Granada, Spain
[84] Universidad de Santiago de Compostela, Spain
[85] Rudolf Peierls Centre for Theoretical Physics, University of Oxford, Oxford, United Kingdom
[87] School of Physics and Astronomy, University of Leeds, United Kingdom
[88] Argonne National Laboratory, Argonne, IL, USA
[89] Case Western Reserve University, Cleveland, OH, USA
[90] Colorado School of Mines, Golden, CO, USA
[91] Colorado State University, Fort Collins, CO, USA
[92] Colorado State University, Pueblo, CO, USA
[93] Fermilab, Batavia, IL, USA
[94] Louisiana State University, Baton Rouge, LA, USA
[95] Michigan Technological University, Houghton, MI, USA
[96] New York University, New York, NY, USA
[97] Northeastern University, Boston, MA, USA
[98] Ohio State University, Columbus, OH, USA
[99] Pennsylvania State University, University Park, PA, USA
[100] Southern University, Baton Rouge, LA, USA
[101] University of Chicago, Enrico Fermi Institute, Chicago, IL, USA
[105] University of Nebraska, Lincoln, NE, USA
[106] University of New Mexico, Albuquerque, NM, USA
[108] University of Wisconsin, Madison, WI, USA
[109] University of Wisconsin, Milwaukee, WI, USA
[110] Institute for Nuclear Science and Technology (INST), Hanoi, Vietnam
[†] Deceased
[a] at Konan University, Kobe, Japan

# The distribution of shower maxima of UHECR air showers

PEDRO FACAL SAN LUIS[1] FOR THE PIERRE AUGER COLLABORATION[2]
[1]*Enrico Fermi Institute & Kavli Institute for Cosmological Physics, University of Chicago, Chicago IL 60637, USA*
[2]*Observatorio Pierre Auger, Av. San Martin Norte 304, 5613 Malargüe, Argentina*
*(Full author list: http://www.auger.org/archive/authors_2011_05.html)*
*auger_spokespersons@fnal.gov*

**Abstract:** We present the measurement of $X_{\max}$, the depth of the maximum of the longitudinal development of ultra high energy air showers, with the fluorescence detector of the Pierre Auger Observatory. After giving an update on the average and fluctuations of $X_{\max}$ with 80% more data than previously published, we discuss the distributions of $X_{\max}$ for different energies and compare it to the predictions of air shower simulations for different primary particles.

**Keywords:** UHECR, The Pierre Auger Observatory, mass composition, shower maxima.

## 1 Introduction

Measuring the cosmic ray composition at the highest energies, along with other measurements such as the flux and the arrival direction distribution, is a key to separate the different scenarios of origin and propagation of cosmic rays. The composition cannot be determined from direct measurements but must be inferred from measurements of the shower that the cosmic ray primary produces in the atmosphere. The atmospheric depth at which this shower attains its maximum size, $X_{\max}$, carries information about the mass of the primary particle and the characteristics of hadronic interactions at very high energy. For a given shower, $X_{\max}$ will be determined by the depth of the first interaction of the primary in the atmosphere, plus the depth that it takes the cascade to develop. The depth of the first interaction is expected to be a decreasing function of the logarithm of the primary energy, while the depth of the shower development rises as $\ln(E)$ [1]. The measured distribution of $X_{\max}$ results from the folding of the distribution of the depth of the first interaction, the shower to shower development fluctuations, and the detector resolution.

The superposition model allows a qualitative treatment of different nuclear primaries of mass $A$: at a given energy $E$, it describes the shower as a superposition of $A$ showers of energy $E/A$. Under this assumption the depth of the maximum of the cascade will be linear with $(\ln(E) - \ln(A))$. Showers of heavier nuclear primaries will develop faster that lighter ones. At the same time the fluctuations of the first interaction will be reduced (by less than $1/\sqrt{A}$ due to correlations between the interactions of the different nucleons). Thus not only the mean value of $X_{\max}$ carries information about the mass of the primary cosmic ray, but the whole distribution is sensitive to the mass composition.

We expect the maximum of the shower to behave as

$$\alpha(\ln E - \ln A) + \beta$$

as function of the energy $E$ and the mass $A$ of the primary. The *elongation rate* is defined as the change of $\langle X_{\max}\rangle$ with energy $D_{10} = \mathrm{d}\langle X_{\max}\rangle/\mathrm{dlog}E$. The parameters $\alpha$ and $\beta$ enclose the dependency of $X_{\max}$ on the properties of the hadronic interactions. There are different theoretical calculations extrapolating the available data to the energies of the interaction between the primary and the atmospheric nucleon [2]. In fact $X_{\max}$ can be used to study the properties of the hadronic interactions at the highest energies [3, 4]. The different hadronic models predict different values for $X_{\max}$, but its dependence on the mass of the primary is qualitatively compatible with the model described here: at a given energy, we expect that for lighter primaries the distribution of $X_{\max}$ will be deeper and broader than the one for heavier primaries.

We use data from the Fluorescence Detector (FD) of the Pierre Auger Observatory [5] to measure the distribution of $X_{\max}$ for ultra high energy cosmic ray showers. First we present an update of the measurements of $\langle X_{\max}\rangle$ and $\mathrm{RMS}(X_{\max})$ as a function of energy with 80% more statistics than previously reported [6]. In addition, we present, for the first time, the measured $X_{\max}$ distributions.

## 2 Data Analysis

Data taken by the Pierre Auger Observatory between December 2004 and September 2010 are used here. The Sur-

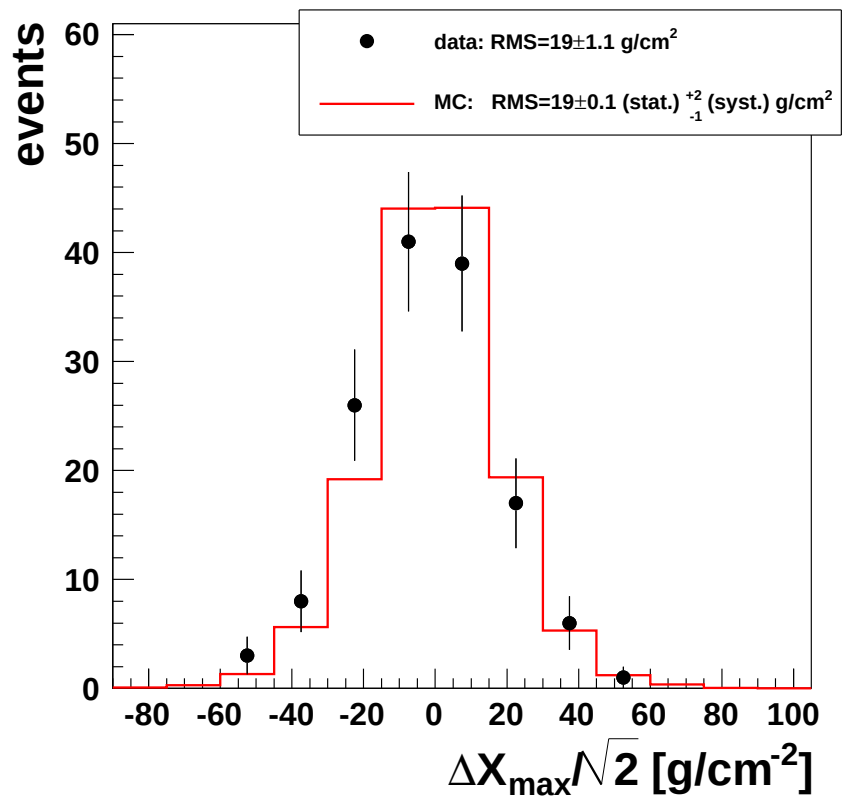

Figure 1: The resolution of $X_{\max}$ obtained using events recorded simultaneously from two FD stations, compared to a detailed Monte Carlo simulation.

face Detector (SD) has 1660 water detector stations arranged in a 1.5 km triangular grid and sensitive to the shower particles at the ground. The FD has 27 telescopes overlooking the SD, housed in 5 different stations, recording UV light emitted in the de-excitation of nitrogen molecules in the atmosphere after the passage of the charged particles of a shower. The shower geometry is reconstructed from the arrival times of the data. The number of fluorescence photons emitted is proportional to the energy deposited in the atmosphere by the shower. Using the shower geometry and correcting for the attenuation of the light between the shower and the detector, the longitudinal profile of the shower can be reconstructed. This profile is fitted to a Gaisser-Hillas function [7] to determine $X_{\max}$ and the energy of the shower [8].

We follow the analysis already reported in [6]. We consider only showers reconstructed using FD data and that have at least a signal in one of the SD stations measured in coincidence. The geometry for these events is determined with an angular uncertainty of 0.6° [9]. The aerosol content in the atmosphere is monitored constantly during data taking [10] and only events for which a reliable measurement of the aerosol optical depth exists are considered. Also the cloud content is monitored nightly across the array and periods with excessive cloud coverage are rejected. Furthermore, we reject events with a $\chi^2/\mathrm{Ndf}$ greater than 2.5 when the profile is fitted to a Gaisser-Hillas, as this could indicate the presence of residual clouds. The total statistical uncertainty in the reconstruction of $X_{\max}$ is calculated including the uncertainties due to the geometry reconstruction and to the atmospheric conditions. Events with uncertainties above 40 $\mathrm{g/cm^2}$ are rejected. We also reject events that have an angle between the shower and the telescope smaller than 20° to account for the difficulties of reconstructing their geometry and for their high fraction of Cherenkov light. Finally, in order to reliably determine $X_{\max}$ we require that the maximum has been actually observed within the field of view of the FD. 15979 events pass this quality selection.

Another set of cuts is used to ensure that the data sample is unbiased with respect to the cosmic ray composition. Since we require data from at least one SD station, we place an energy dependent cut on both the shower zenith angle and the distance of the SD station to the reconstructed core so the trigger probability of a single station at these energies is saturated for both proton and iron primaries.

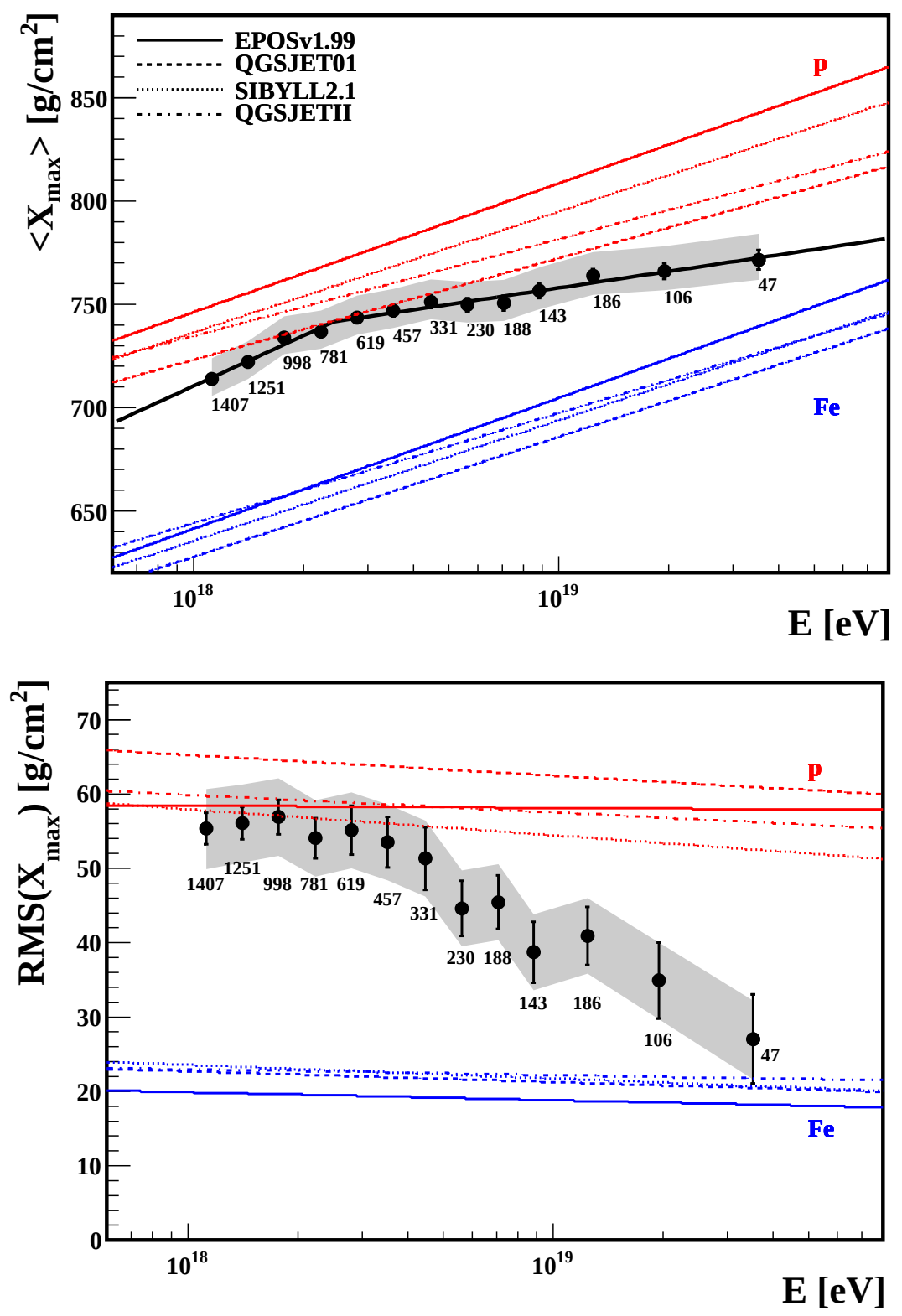

Figure 2: $\langle X_{\max}\rangle$ (top panel) and $\mathrm{RMS}\,(X_{\max})$ (bottom panel) as a function of the energy. Data (points) are shown with the predictions for proton and iron for several hadronic interaction models. The number of events in each bin is indicated. Systematic uncertainties are indicated as a band.

Finally, requiring that the shower maximum is observed means that, for some shower geometries, we could introduce a composition dependent bias in our data. This is avoided using only geometries for which we are able to observe the full range of the $X_{\max}$ distribution.

At the end 6744 events (42% of those that pass the quality cuts) remain above $10^{18}$ eV. The systematic uncertainty in the energy reconstruction of the FD events is 22% The resolution in $X_{\max}$ is at the level of 20 $\mathrm{g/cm^2}$ over the energy range considered. This resolution is estimated with a detailed simulation of the detector and cross-checked using the difference in the reconstructed $X_{\max}$ when one event is observed by two or more FD stations (Fig. 1).

## 3 Results and discussion

In Fig. 2 we present the updated results for $\langle X_{\max}\rangle$ and $\mathrm{RMS}\,(X_{\max})$ using 13 bins of $\Delta\log E = 0.1$ below $10^{19}$ eV and $\Delta\log E = 0.2$ above. An energy dependent correction ranging from 3.5 $\mathrm{g/cm^2}$ (at $10^{18}$ eV) to $-0.3\ \mathrm{g/cm^2}$ (at $7.2\cdot10^{19}$ eV, the highest energy event) has been applied to the data to correct for a small bias observed when reconstructing Monte Carlo simulated events. The total systematic uncertainty in $\langle X_{\rm max}\rangle$ goes from 10 ${\rm g/cm^2}$ at low energy to 13 ${\rm g/cm^2}$ at high energy. It includes contributions from the uncertainties in the calibration, the atmospheric data,the reconstruction and the event selection.

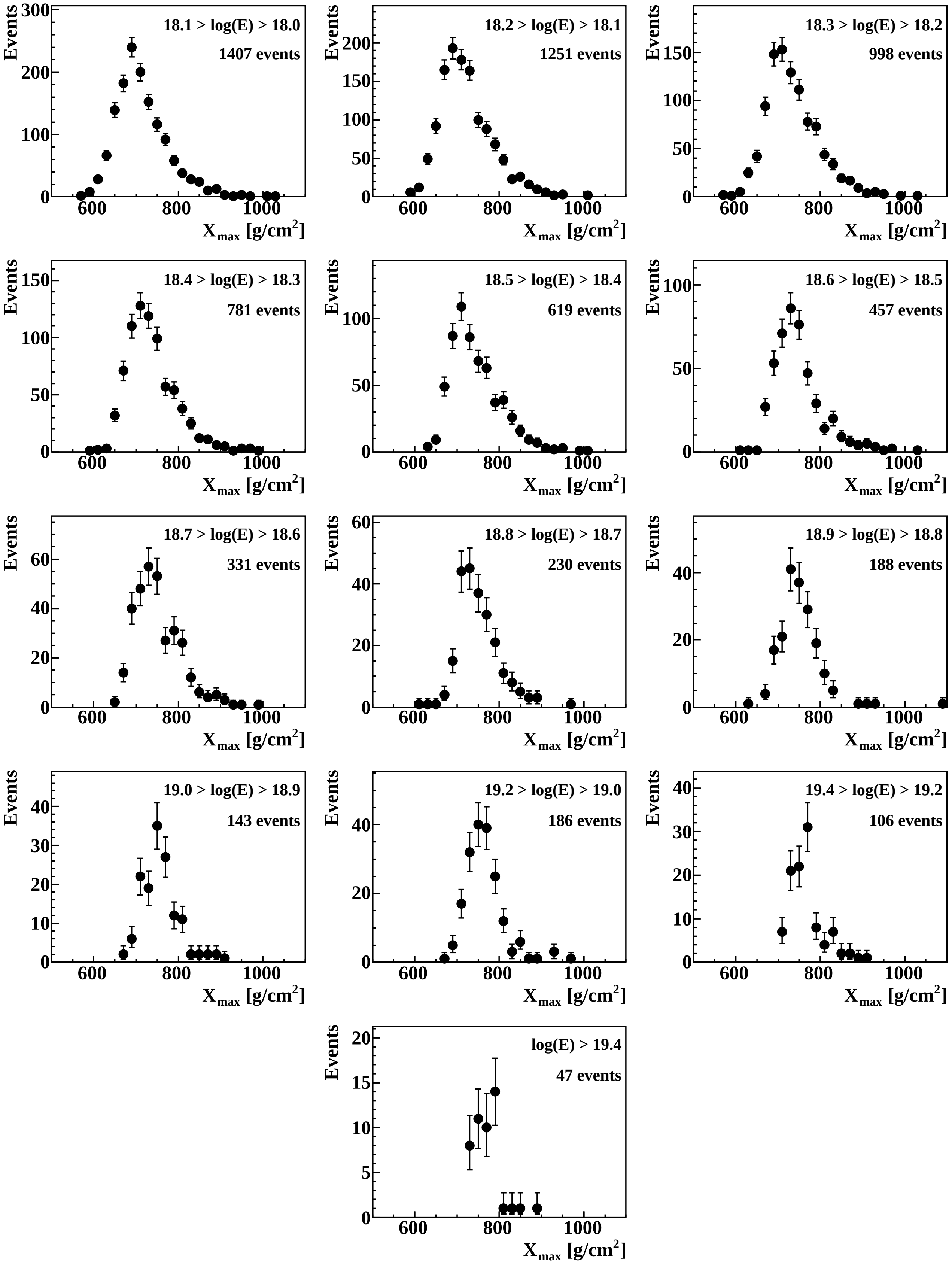

Figure 3: Distribution of $X_{\max}$. The values of the energy limits and the number of events selected are indicated for each panel.

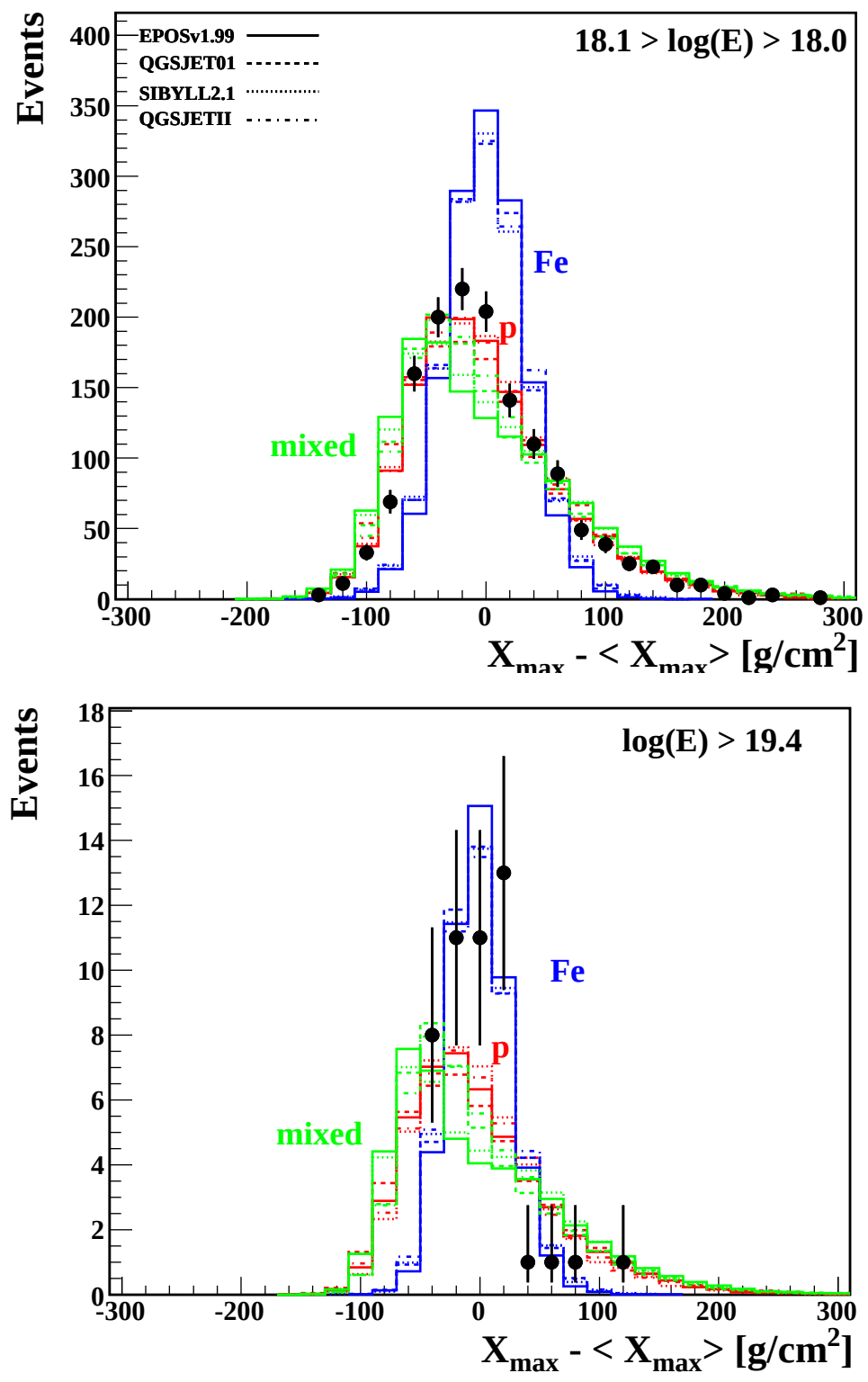

Figure 4: Centered distribution, $X_{\rm max} - \langle X_{\rm max}\rangle$, for the lowest and highest energy bins. Subtraction of the mean allows only for the comparison of the shapes of these distributions with the superimposed MC simulations (see text). Mixed is 50% p and Fe.

The elongation rate is best described using two slopes. Below $\log(E/{\rm eV}) = 18.38^{+0.07}_{-0.17}$ we obtain $D_{10} = 82^{+48}_{-\ 8}$ ${\rm g/cm^2/decade}$ and above $D_{10} = 27^{+3}_{-8}$ ${\rm g/cm^2/decade}$, with a $\chi^2/{\rm Ndf} = 7.4/9$. A fit using one slope does not describe our data well ($\chi^2/{\rm Ndf} = 54/11$). The small elongation rate at high energies could be interpreted as a change in composition of cosmic rays, from lighter primaries to heavy. Due to the small energy range, the low energy elongation rate has large statistical uncertainties (an extension of this analysis towards lower energy will soon be possible using the data of the recently installed high elevation telescopes [11]). The results of the fit are fully compatible with those in [6].

The ${\rm RMS}\,(X_{\rm max})$ reported in Fig. 2 has been corrected subtracting in quadrature a resolution that goes from 27 ${\rm g/cm^2}$ at low energy to 18 ${\rm g/cm^2}$ at high energy. The systematic uncertainty on ${\rm RMS}\,(X_{\rm max})$ is at the level of 5 ${\rm g/cm^2}$. ${\rm RMS}\,(X_{\rm max})$ decreases gradually with energy, from 55 ${\rm g/cm^2}$ to 26 ${\rm g/cm^2}$. The decrease with the energy becomes steeper around the same point where the two sections of the $\langle X_{\rm max}\rangle$ fit are joined.

$\langle X_{\rm max}\rangle$ and ${\rm RMS}\,(X_{\rm max})$ have been obtained from the distributions shown in Fig. 3. The reduction in the width of the distribution as the energy increases can be clearly observed from the figures. This reduction is even more striking for the tail of the distribution towards deep $X_{\rm max}$: the proton-like tail at low energy gives gradually way to much more symmetric distributions with smaller tails.

For most of the models, the data would have to be adjusted within their systematic uncertainties to simultaneously match both $\langle X_{\rm max}\rangle$ and ${\rm RMS}\,(X_{\rm max})$ to a given composition mixture ($\langle X_{\rm max}\rangle$ downward and ${\rm RMS}\,(X_{\rm max})$ upward and/or the energy scale upward). As can be seen in Fig. 2, the MC predictions are more uncertain for the $\langle X_{\rm max}\rangle$ than for the fluctuations. This is mainly due to the additional dependence of $\langle X_{\rm max}\rangle$ on the multiplicity in hadronic interactions [3]. In Fig. 4 we therefore compare the *shape* of the distributions, $X_{\rm max} - \langle X_{\rm max}\rangle$, to MC predictions for different compositions and hadronic interaction models. As can be seen, in this representation the various models predict a nearly universal shape. At low energy, the shape of the data is compatible with a very light or mixed composition, whereas at high energies, the narrow shape would favour a significant fraction of nuclei (CNO or heavier). It is, however, worthwhile noting, that both the mixed composition and the pure iron predictions are at odds with the measured $\langle X_{\rm max}\rangle$. Also, a significant departure from the predictions of the available hadronic models would modify this interpretation (see [4] for an estimate of the properties of hadronic interactions up to $10^{18.5}$ eV using these data and [12] for a comparison between our data and some of the model predictions).

## References

[1] J. Matthews, Astropart. Phys., 2005, **22**: 387-397.
[2] N.N. Kalmykov and S.S. Ostapchenko, Phys. Atom. Nucl., 1993, **56**: 346-353; S.S. Ostapchenko, Nucl. Phys. B (Proc. Suppl.), 2006, **151**: 143-146; T. Pierog and K. Werner, Phys. Rev. Lett., 2008, **101**: 171101. E.-J. Ahn *et al.*, Phys. Rev., 2009, **D80** : 094003.
[3] R. Ulrich *et al.*, Phys. Rev., 2011, **D83** : 054026.
[4] R. Ulrich, for The Pierre Auger Collaboration, paper 0946, these proceedings.
[5] The Pierre Auger Collaboration, Nucl. Instrum. Meth., 2010, **A620**: 227-251.
[6] The Pierre Auger Collaboration, Phys. Rev. Lett., 2010, **104**: 091101.
[7] T.K. Gaisser and A.M. Hillas, Proc. 15th ICRC, 1977, **8**: 353-357.
[8] M. Unger *et al.*, Nucl. Instrum. Meth., 2008, **A588**: 433-441.
[9] C. Bonifazi, for The Pierre Auger Collaboration, Nucl. Phys. B (Proc. Suppl.), 2009, **190**: 20-25.
[10] The Pierre Auger Collaboration, Astropart. Phys., 2010, **33** : 108-129.
[11] H.-J. Mathes, for The Pierre Auger Collaboration, paper 0761, these proceedings.
[12] J. Allen, for The Pierre Auger Collaboration, paper 0703, these proceedings.

# Measurement of Atmospheric Production Depths of muons with the Pierre Auger Observatory

D. GARCÍA-GÁMEZ[1] FOR THE PIERRE AUGER COLLABORATION[2]
[1]*Laboratoire de l'Accélérateur linéaire (LAL), Université Paris 11, CNRS-IN2P3, Orsay, France*
[2]*Observatorio Pierre Auger, Av. San Martin Norte 304, 5613 Malargüe, Argentina*
*(Full author list: http://www.auger.org/archive/authors_2011_05.html)*
*auger_spokespersons@fnal.gov*

**Abstract:** The surface detector array of the Pierre Auger Observatory provides information about the longitudinal development of the hadronic component of extensive air showers in an indirect way. In this contribution we show that it is possible to reconstruct the Muon Production Depth distribution (MPD) using the FADC traces of surface detectors far from the shower core. We characterize the goodness of this reconstruction for zenith angles around 60° and different energies of the primary particle. From the MPDs we define $X^{\mu}_{\max}$ as the depth, along the shower axis, where the number of muons produced reaches a maximum. We explore the potentiality of $X^{\mu}_{\max}$ as a sensitive parameter to determine the mass composition of cosmic rays.

**Keywords:** Muon Production Depth distributions, Pierre Auger Observatory

## 1 Introduction

The Pierre Auger Observatory was conceived to study the properties of Ultra-High Energy Cosmic Rays (UHECR). It is a hybrid detector that combines both surface and fluorescence detectors at the same site [1]. The origin and chemical composition of UHECR are still an enigma. Currently, the most sensitive parameter to analyse mass composition is the depth of the shower maximum, $X_{\max}$, see e.g. [2, 3], measured by the fluorescence detector (FD) [4]. The fluorescence detector operates only on clear, moonless nights, so its duty cycle is small (about 13 %). On the other hand, the surface detector array (SD) [5] has a duty cycle close to 100 %. This increase in statistics makes any SD-based observable of great interest to study the composition of UHECR.

In an extensive air shower (EAS) muons are mainly produced by the decay of pions and kaons. Their production points are constrained to a region very close to the shower axis, of the order of tens of meters [6]. Muons can be taken as travelling along straight lines to ground, due to the lesser importance of bremsstrahlung and multiple scattering effects compared to other geometrical and kinematical factors. In [6, 7] these features are exploited to build a model for obtaining the muon production depth (MPD) along the shower axis. The MPDs are calculated from the muon time structure at ground. These times are given along with the times of the other particles reaching ground by the FADCs of the SD. In this work we show that MPDs provide a physical observable that can be used as a sensitive parameter to study the chemical composition of cosmic rays [8].

## 2 MPD reconstruction

Starting from the time signals that muons produce in the surface detectors, the model discussed in [6, 7] derives from geometrical arguments the distribution of muon production distance, $z$:

$$z = \frac{1}{2}\left(\frac{r^2}{ct_g} - ct_g\right) + \Delta \tag{1}$$

where $r$ is the distance from the point at ground to the shower axis, $\Delta$ is the distance from the same point to the shower plane and $t_g$ (*geometrical delay*) is the time delay with respect to the shower front plane. The shower front plane is defined as the plane perpendicular to the shower axis and moving at the speed of light, $c$, in the direction of the shower axis. It contains the first interaction point and also the core hitting ground. This calculation assumes that muons travel at the speed of light. If we account for their finite energy $E$, the total time delay would be $ct = ct_g + ct_\epsilon(E)$. This extra contribution is dominant at short distances to the core, where the geometrical time delay is very small. At large distances ($r > 600$ m) the *kinematic delay*, $t_\epsilon$, acts as a correction (typically below 20%). It must be subtracted from the measured time delay prior to the conversion into $z$, as described in [6, 7].

Equation 1 gives a mapping between the production distance $z$ and the *geometrical delay* $t_g$ for each point at

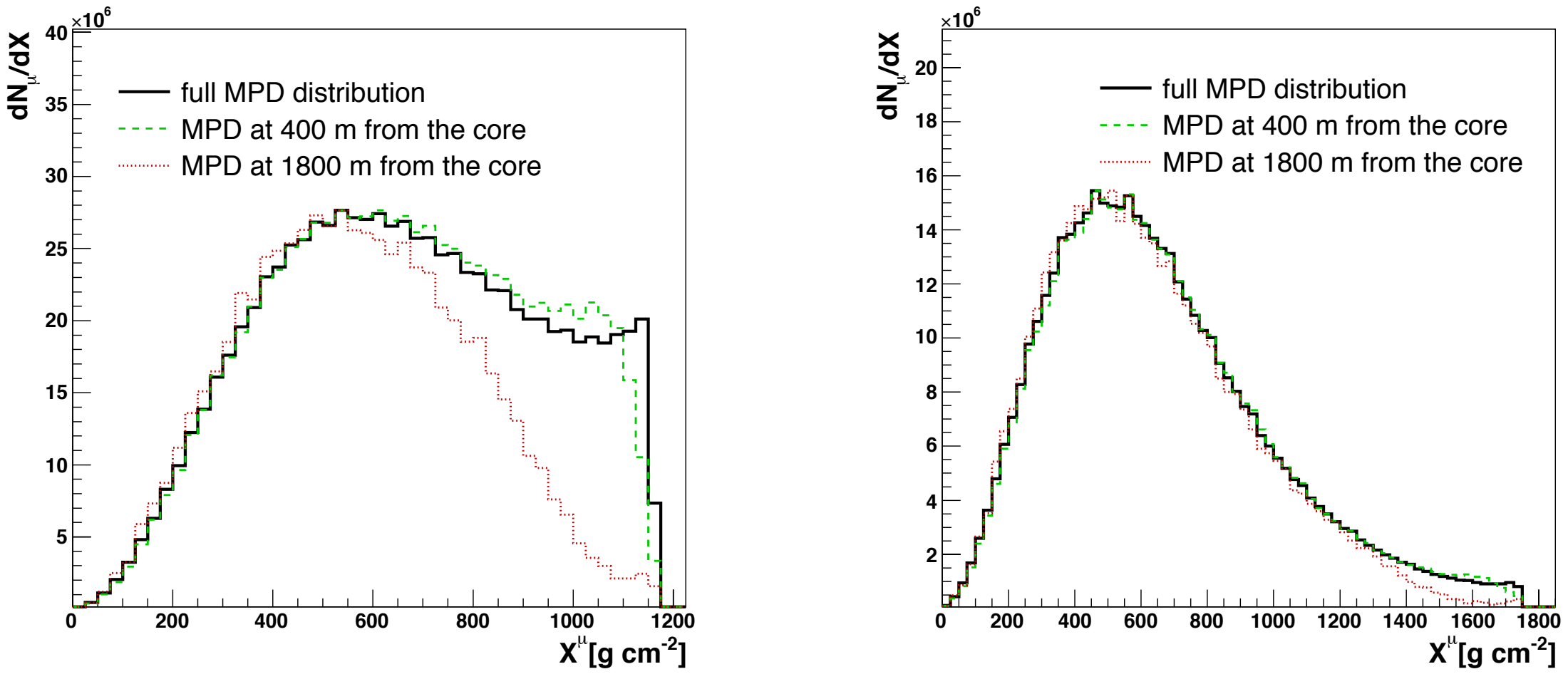

Figure 1: Muon production depth distributions (MPDs) extracted from an iron shower of $10^{19}$ eV simulated with AIRES [10] at two different zenith angles: 41° (left) and 60° (right). The MPD dependence with the distance to the core is shown.

ground. The production distance can be easily related to the total amount of traversed matter $X^\mu$ using

$$X^\mu = \int_z^\infty \rho(z')dz' \qquad (2)$$

where $\rho$ is the atmosphere density. This $X^\mu$ distribution is referred to as MPD. The shape of the MPD contains relevant information about the development of the hadronic cascade and the first interaction point. To extract valuable physics insight from the MPD we perform a fit. It was found that a Gaisser-Hillas function [9] can describe the shape of the MPD well. The fit with this function provides the maximum of the distribution, $X^\mu_{\rm max}$. We interpret $X^\mu_{\rm max}$ as the point where the production of muons reaches the maximum along the cascade development. As shown in the following sections, this new observable can be used for composition studies.

The MPD is populated with the surviving muons reaching ground, so its shape depends on the zenith angle. Figure 1 displays MPDs directly extracted from AIRES simulations [10] at different zenith angles and at different distances from the core, $r$. For angles of about 40° and lower, the shape of the MPD and the position of its maximum show a strong $r$ dependence. However, at zenith angles of around 60° and above, where the showers develop very high in the atmosphere, the differences between the MPD at different distances to the core become small. Thus, for those showers we can add in the same histogram the $X^\mu$ values given by the time signals from the different surface detectors. The addition of the signals from the different surface detectors contributing to the MPD at small zenith angles would demand the introduction of a correction factor that transforms all those signals to the one expected at a reference $r$ (see [6, 7] for a thorough discussion about this correction). At larger zenith angles the distortion due to the detector time resolution becomes larger. The above reasons lead us to select the data with measured zenith angles between 55° and 65° for our analysis.

## 2.1 Detector effects

The precision of the method is limited so far by the detector capabilities. The total uncertainty of the MPD maximum, $\delta X^\mu_{\rm max}$, decreases as the square root of the number of muons $N_\mu$, and decreases quadratically with the distance to the core $r$. This last uncertainty is linked to each single time bin entry of the FADC traces. To keep the distortions on the reconstructed MPD small, only detectors far from the core can be used. The cut in $r$ diminishes the efficiency of the reconstruction, as the number of muons contributing to the MPD is reduced. Hence a $r_{cut}$ value must be carefully chosen in order to guarantee good reconstruction efficiencies, avoiding at the same time a bias on the mass of the primary.

Furthermore, signals collected by the water Cherenkov detectors are the sum of the electromagnetic (EM) and muonic components. Both exhibit a different arrival time behavior. As a consequence, a cut on signal threshold, rejecting all time bins with signal below a certain value, might help diminishing the contribution of the EM contamination. The so called EM halo, coming from the decay of muons in flight, is harder to suppress. But this component follows closer the time distribution of their parent muons, thus it does not hamper our analysis.

## 2.2 Reconstruction cuts

To study and select the cuts needed for a good MPD reconstruction and an accurate $X^\mu_{\rm max}$ determination we have used Monte Carlo simulations. The selection of cuts must be a trade off between the resolution of the reconstructed MPD and the number of muons being accepted into such reconstruction. The chosen $r_{cut}$ is energy independent. This implies that any difference in resolution that we find for different energies will be mainly a consequence of the different amount of muons detected at ground. In our analysis, we consider only those detectors whose distance to the shower core is larger than 1800 m. To reduce residual EM contamination and potential baseline fluctuations we have applied a mild cut on the threshold of the FADC signals used to build the MPD. We have discarded FADC bins where the signal is below 0.3 VEM. Finally, the MPD is reconstructed adding those detectors whose total recorded signal is above 3 VEM. This requirement is set to avoid, in real data, the contribution of detectors (usually far away from the core) having a signal dominated by accidental particles.

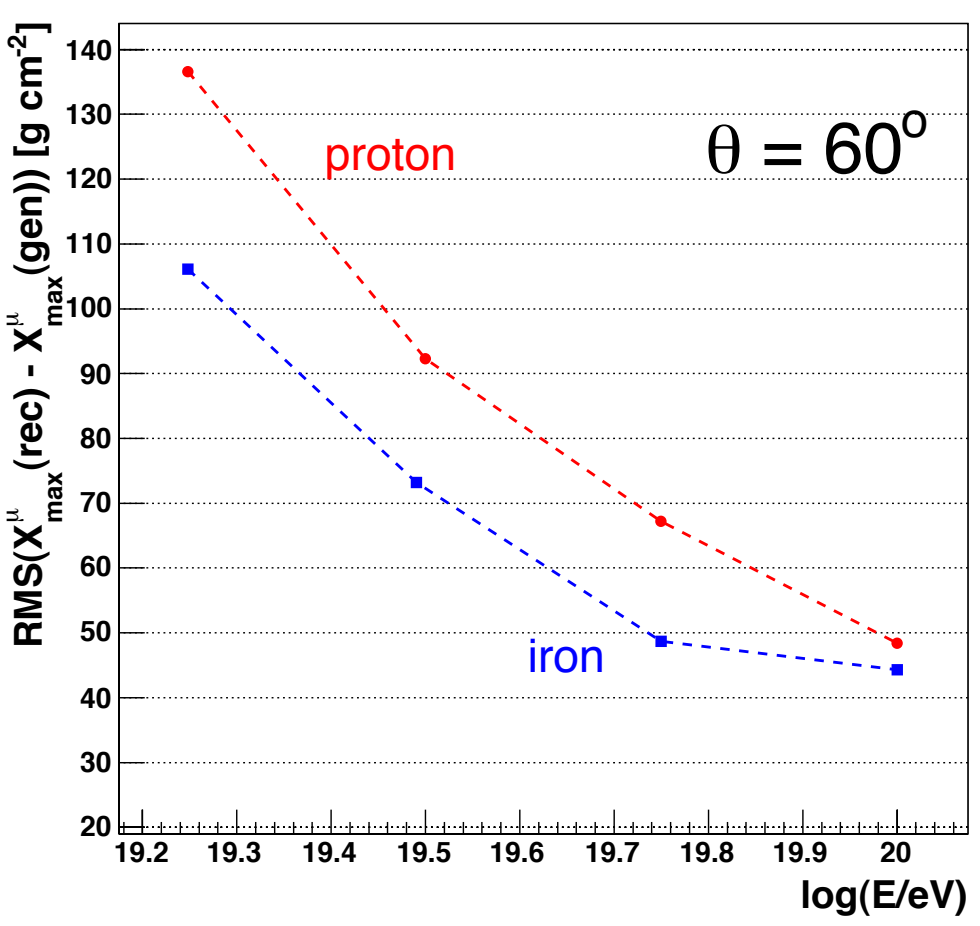

Figure 2: Energy evolution of the resolution we obtain, on an event by event basis, when we reconstruct $X^{\mu}_{\max}$ for showers generated with AIRES and QGSJETII [11].

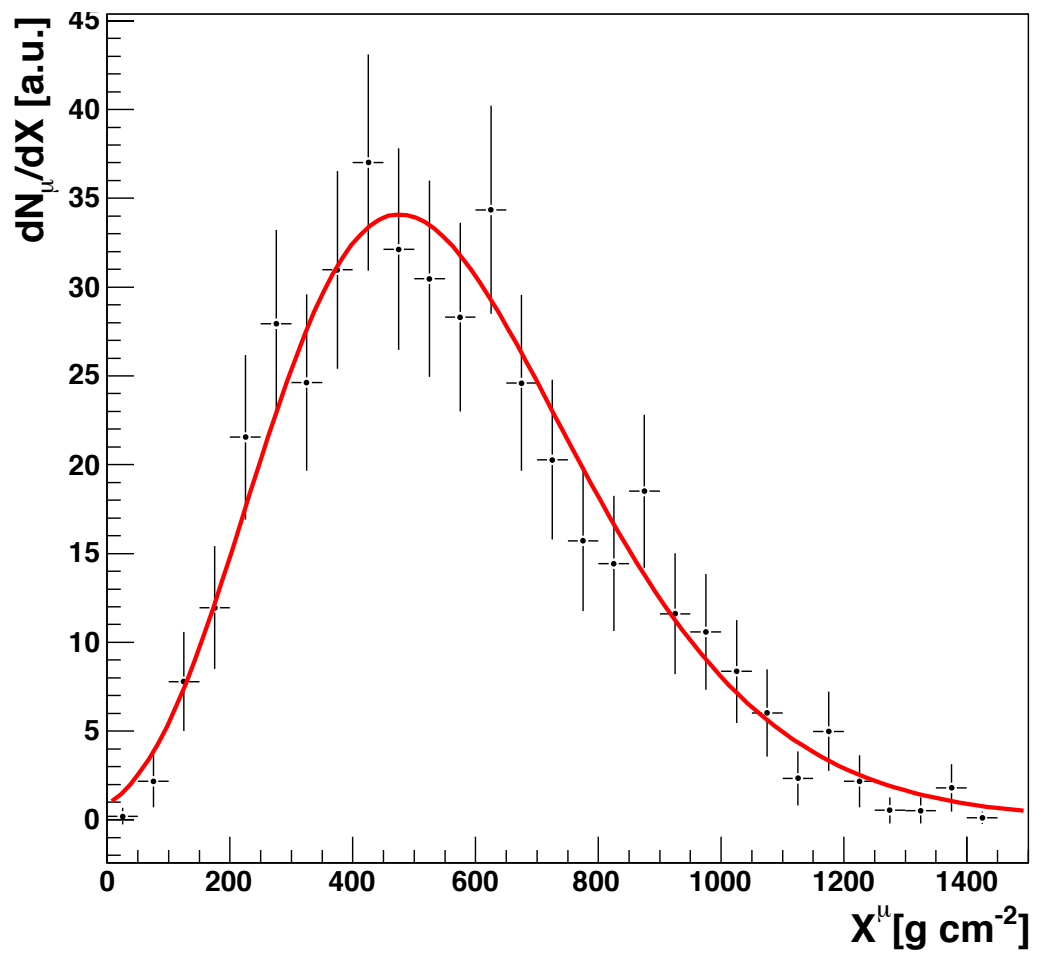

Figure 3: Real reconstructed MPD, $\theta$ = (59.05 $\pm$ 0.07) $^\circ$ and E = (94 $\pm$ 3) EeV, with its fit to a Gaisser-Hillas function.

This set of cuts has a high muon selection efficiency. Regardless of the energy of the primary and its composition, muon fractions above 85% are always obtained. This guarantees an EM contamination low enough to obtain an accurate value of $X^{\mu}_{\max}$.

### 2.3 Selection cuts

To optimize the quality of our reconstructed profiles we apply the following cuts:

- **Trigger cut**: All events must fulfill the T5 trigger condition [5].
- **Energy cut**: Since the number of muons is energy dependent, $N_{\mu} \propto E^{\alpha}/r^{\beta}$, we have observed that in events with energies below 20 EeV the population of the MPD is very small, giving a very poor determination of the $X^{\mu}_{\max}$ observable. Therefore we restrict our analysis to events with energy larger than 20 EeV.
- **Fit quality**: Only events with a good MPD fit ($\chi^2$/ndf $< 2.5$) to a Gaisser-Hillas function are accepted.
- **Shape cut**: The reduced $\chi^2$ of a straight line and a Gaisser-Hillas fit must satisfy $\chi^2_{GH}$/ndf $<$ $2\chi^2_{line}$/ndf.
- **Curvature**: When the fitted radius of curvature of the shower front, $R$, is very large we observe an underestimation of the reconstructed $X^{\mu}_{\max}$. So only events with $R <$ 29000 m are included in our analysis.

The overall event selection efficiencies are high ($>$ 80%) and the difference between iron and proton is small for the whole range of considered energies (see Table 1). Our cuts do not introduce any appreciable composition bias. We finally note that for the set of surviving events, the bias in the $X^{\mu}_{\max}$ reconstruction is between $\pm$ 10 g cm$^{-2}$, regardless of the initial energy or the chemical composition of the primary. The resolution ranges from about 120 g cm$^{-2}$ at the lower energies to less than 50 g cm$^{-2}$ at the highest energy (see Figure 2).

We note that the predictions of $X^{\mu}_{\max}$ from different hadronic models (such as those shown in Figure 4) would not be affected if a discrepancy between a model and data [12] is limited to the *total* number of muons. However, differences in the muon energy and spatial distribution would modify the predictions.

## 3 Application to real data

Our analysis makes use of the data collected between January 2004 and December 2010. Our initial sample of events with zenith angle $\theta \in [55^\circ, 65^\circ]$ and a reconstructed energy bigger than 20 EeV consists of 417 events. The overall selection efficiency amounts to 58%, which translates into 244 surviving events. The difference between the efficiencies shown in Table 1 and the selection efficiency in real data is due to the T5 cut [5]. This cut has an efficiency of about 72% for data, while all our Monte Carlo showers are generated as T5 events. We compute MPDs on an event by event basis. Figure 3 shows the reconstructed MPD for one of our most energetic events. The evolution of the $\langle X^{\mu}_{\max}\rangle$ observable as a function of energy is shown in Figure 4. The selected data has been grouped into five bins of energy. Each bin has a width of 0.1 in $\log_{10}$(E/eV), except the last one which contains all the events with energy larger than $\log_{10}$(E/eV)=19.7. The error bars correspond to the ratio between the RMS of the distributions of $X^{\mu}_{\max}$ and the square root of the number of entries. If compared to air shower predictions using standard interaction models, our measurement is compatible with a mixed composition.

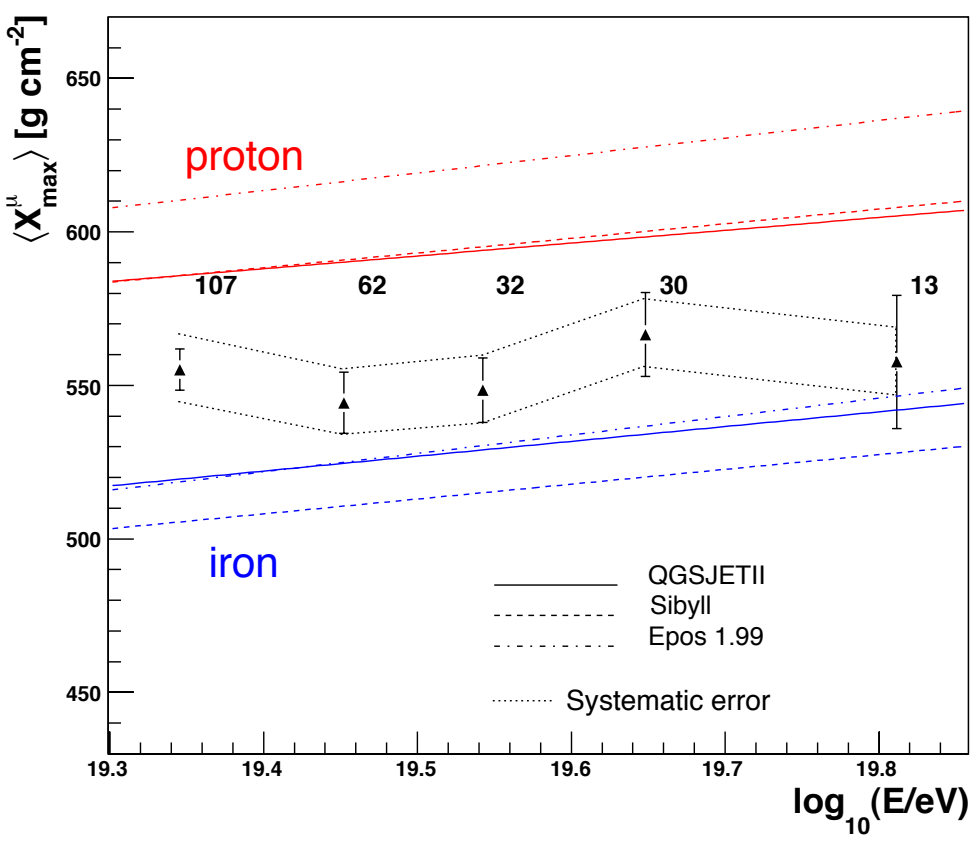

Figure 4: $\langle X^{\mu}_{\max}\rangle$ as a function of the energy. The number of real data events in each energy bin is indicated. The predictions for proton and iron following different hadronic models are shown as well.

Table 1: Selection efficiencies for proton and iron QGSJETII Monte Carlo showers as a function of energy.

| $\log_{10}$(E/eV) | $\varepsilon_p$ (%) | $\varepsilon_{Fe}$ (%) | $\lvert\varepsilon_p - \varepsilon_{Fe}\rvert$ (%) |
|---|---|---|---|
| 19.25 | 82 | 87 | 6 |
| 19.50 | 84 | 86 | 2 |
| 19.75 | 85 | 82 | 3 |
| 20.00 | 95 | 97 | 2 |

Table 2 lists the most relevant sources contributing to the systematic uncertainty. The uncertainties on the MPD reconstruction and event selection translate into a systematic uncertainty on $\langle X^{\mu}_{\max}\rangle$ of 11 g cm$^{-2}$.

## 4 Conclusions

We have shown that it is possible to reconstruct the muon production depth distribution using the FADC traces of the SD detectors far from the core. From the MPDs we define a new observable $X^{\mu}_{\max}$. It measures the depth along the shower axis where the number of produced muons reaches a maximum. We have characterized the applicability of this observable and analysed its resolution for zenith angles $\sim 60^\circ$ and different shower energies. We have demonstrated, for the first time, that $X^{\mu}_{\max}$ is a parameter sensitive to the mass composition of UHECR. The result of this study is in agreement with all previous Auger results [13] obtained with other completely independent methods.

Table 2: Evaluation of the main sources of systematic uncertainties.

| Source | Sys. Uncertainty (g cm$^{-2}$) |
|---|---|
| Reconstruction bias | 9.8 |
| Core position | 4.8 |
| EM contamination | 1.5 |
| $\chi^2$ cut | 0.2 |
| Selection efficiency | 1 |
| Total | 11 |

## References

[1] J. Abraham *et al.* , Nucl. Instr. and Meth. A, 2004, **523**(1-2): 50-95
[2] J. Abraham *et al.* , Phys. Rev. Lett., 2010, **104** (9): 1-7
[3] R. U. Abbasi *et al.* , Phys. Rev. Lett., 2010, **104**(16): 161101
[4] J. Abraham *et al.* , Nucl. Instr. and Meth. A, 2010, **620**(2-3): 227-251
[5] J. Abraham *et al.* , Nucl. Instr. and Meth. A, 2010, **613**(1): 29-39
[6] L. Cazon, R.A. Vazquez, A.A. Watson, E. Zas, Astropart. Phys., 2004, **21**(1): 71-86
[7] L. Cazon, R.A. Vazquez, E. Zas, Astropart. Phys., 2005, **23**(4): 393-409
[8] D. Garcia-Gamez, Muon Arrival Time distributions and its relationship to the mass composition of Ultra High Energy Cosmic Rays: An application to the Pierre Auger Observatory, Universidad de Granada, PhD Thesis, 2010
[9] T. K. Gaisser and A. M. Hillas, Proceedings, 15th International Cosmic Ray Conference, **8** , 13-26
[10] S. Sciutto, Proceedings, 27th International Cosmic Ray Conference, 2001, arXiv: astro-ph/0106044
[11] S. Ostapchenko, AIP Conference Proceeding, 2007, **928**, 118-125
[12] J. Allen, for the Pierre Auger Collaboration, paper 0703, these proceedings
[13] D. Garcia-Pinto, for the Pierre Auger Collaboration, paper 0709, these proceedings

# Measurements of the Longitudinal Development of Air Showers with the Pierre Auger Observatory

DIEGO GARCIA-PINTO[1] FOR THE PIERRE AUGER COLLABORATION[2]
[1]*Universidad Complutense de Madrid, Madrid, Spain*
[2]*Observatorio Pierre Auger, Av. San Martin Norte 304, 5613 Malargüe, Argentina*
*(Full author list: http://www.auger.org/archive/authors_2011_05.html)*
*auger_spokespersons@fnal.gov*

**Abstract:** Due to its hybrid design, the Pierre Auger Observatory provides a variety of independent experimental observables that carry information on the characteristics of the longitudinal development of ultra-high energy air showers. These include the direct measurement of the profile of the energy deposit of showers in the atmosphere through the detection of fluorescence light but also observables derived from the shower signal measured with the surface detector array. In this contribution we present a comparison of the results obtained with the fluorescence detector on the depth of shower maximum with complementary information derived from asymmetry properties of the particle signal in the surface detector stations and the depth profile of muon production points, also derived from surface detector data. The measurements are compared to predictions for proton- and iron-induced showers.

**Keywords:** UHECR, The Pierre Auger Observatory, mass composition, hadronic interactions

## 1 Introduction

The properties of ultra-high energy cosmic rays (UHECRs) can be studied by measuring the extensive air showers (EAS) that they produce in the atmosphere. For example, information on the mass of the primary particles can, in principle, be derived from the longitudinal depth profiles of these showers. However, the longitudinal development of the showers are strongly affected by the mass composition of cosmic rays and by the features of the hadronic interactions, both of which vary with energy in a manner that is unknown. If one were confident about the behaviour of one of these quantities then the behaviour of the other could be deduced.

In this article we present the measurement of four independent observables that are closely related to the longitudinal depth profile of air showers and hence, sensitive to primary mass composition. Due to the different character of the observables employed, a direct comparison of the measurement results is not possible. Instead, the data are compared to predictions from air shower simulations. Modelling uncertainties are considered by using the three different interaction models EPOS, QGSJET II, and SIBYLL [1], but it is understood that the differences between these models might not fully represent the theoretical uncertainties [2].

## 2 Measurements of the Longitudinal Shower Development

With the Pierre Auger Observatory [3] information on the shower development can be extracted using both the Surface Detector (SD) and the Fluorescence Detector (FD). The SD consists of more than 1660 detector stations covering an area of approximately $3000\,\mathrm{km}^2$. Each SD unit is a water-Cherenkov detector with electronics that digitises the signal at 40 MHz sampling rate. The FD has a total of 27 optical telescopes arranged in five sites overseeing the SD.

The observation of showers with the FD allows us to directly measure the most important observable to characterise the longitudinal profile of a shower, the depth of the shower maximum, $X_{\max}$, i.e. the depth at which air showers deposit the maximum energy per unit mass of atmosphere traversed [4]. On the other hand, the SD provides observables which are related to the longitudinal shower profile as well. These observables are subject to independent systematic uncertainties (both experimentally and theoretically). Moreover the higher statistics of showers measured with the SD allows us to reach higher energies than with the FD.

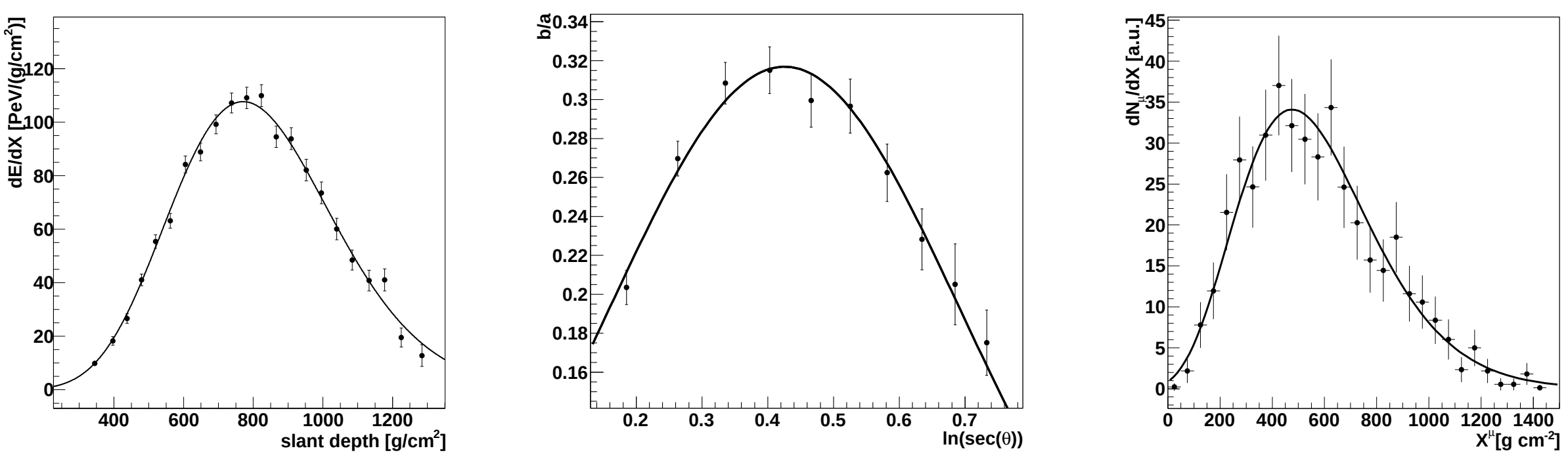

Figure 1: Typical longitudinal development of the energy deposit (left panel), of the average asymmetry in the risetime (centre panel) and the muon production depth (right panel).

## 2.1 Depth of Shower Maximum

The measurement of the longitudinal profile of the energy deposit in the atmosphere with the Pierre Auger Observatory is described in [4]. In this analysis hybrid events, i.e. showers observed simultaneously by the FD and at least by one SD station, have been used. The longitudinal profile of the energy deposit is reconstructed by the FD from the recorded fluorescence and Cherenkov light signals. The collected light is corrected for the attenuation between the shower and the detector using data from atmospheric monitoring devices. The longitudinal shower profile is reconstructed as a function of the atmospheric depth and $X_{\rm max}$ is obtained by fitting the profile with a Gaisser-Hillas function. A typical longitudinal profile of the energy deposit of one shower is shown in the left panel of Fig. 1.

The $X_{\rm max}$ results presented here are an update of [4]. Hybrid events recorded between December 2004 and September 2010 with reconstructed energy above $10^{18}$ eV have been used for the present analysis. To obtain a good resolution in the measurement of $X_{\rm max}$, several quality cuts are applied. The cuts and their effects are described fully in [5]. After all cuts, 6744 events are selected for the $X_{\rm max}$ analysis. The average values of the shower maximum, $\langle X_{\rm max}\rangle$, as a function of energy are displayed in Fig. 2, alongside predictions from several models. Uncertainties of the atmospheric conditions, calibration, event selection and reconstruction give rise to a systematic uncertainty of $\leq 13\ {\rm g/cm^2}$ [4] which corresponds to $\lesssim 13$ % of the proton-iron separation predicted by the models. Since the $X_{\rm max}$ resolution of the FD is at the level of 20 ${\rm g/cm^2}$ above a few EeV, the intrinsic shower-to-shower fluctuations, RMS($X_{\rm max}$), can be measured as well, see lower panel of Fig. 2.

## 2.2 Asymmetry of Signal Risetime

For each SD event, the water-Cherenkov detectors record the signal as a function of time. The first part of the signal is dominated by the muon component which arrives earlier and over a period of time shorter than the electromagnetic particles, since muons travel in almost straight lines whereas the electromagnetic particles suffer more multiple scattering on their way to ground. Due to the absorption of the electromagnetic (EM) component, the number of these particles at the ground depends, for a given energy, on the distance to the shower maximum and therefore on the primary mass. In consequence, the time profile of particles reaching ground is sensitive to cascade development as the higher is the production height the narrower is the time pulse.

The time distribution of the SD signal is characterised by means of the risetime (the time to go from 10% to 50% of the total integrated signal), $t_{1/2}$, which depends on the distance to the shower maximum, the zenith angle $\theta$ and the distance to the core $r$. In previous studies [6] the risetime was related to the shower maximum using a subset of hybrid events. Using this correlation it is possible to measure the shower evolution with surface detector data.

The azimuthal asymmetry of $t_{1/2}$ from water-Cherenkov detector signals of non-vertical showers carries information about the longitudinal development of the showers [7]. Unfortunately it is not possible to define the asymmetry on an event-by-event basis, therefore the risetime asymmetry is obtained by grouping events in bins of energy and $\sec\theta$. A key parameter for the analysis is the angle $\zeta$, the azimuth angle in the shower plane (the plane perpendicular to the shower axis). Detectors that are struck early in the development of the shower across the array have values of this angle in the range $-\pi/2 < \zeta < \pi/2$ with $\zeta = 0°$ corresponding to the vertical projection of the incoming direction onto the shower plane. For each ($E$, $\sec\theta$) bin a fit of $\langle t_{1/2}/r\rangle = a + b\cos\zeta$ provides the asymmetry amplitude, $b/a$. For a given energy, the $b/a$ value changes with the zenith angle, i.e. distance to the shower maximum. The evolution of $b/a$ with zenith angle is thus reminiscent of the longitudinal development of the shower and has a maximum which is different for different primaries [8]. For each energy bin, the asymmetry amplitude is fitted using a Gaussian function of $\ln(\sec\theta)$. This allows the determination of the position of the maximum, $\Theta_{\rm max}$, defined as the value of $\sec\theta$ for which $b/a$ is maximum. In Fig.1, centre panel, an example of $b/a$ as a function of $\ln(\sec\theta)$ and the corresponding fit to obtain $\Theta_{\rm max}$ is shown for the energy bin of $\log(E/{\rm eV}) = 18.85 - 19.00$.

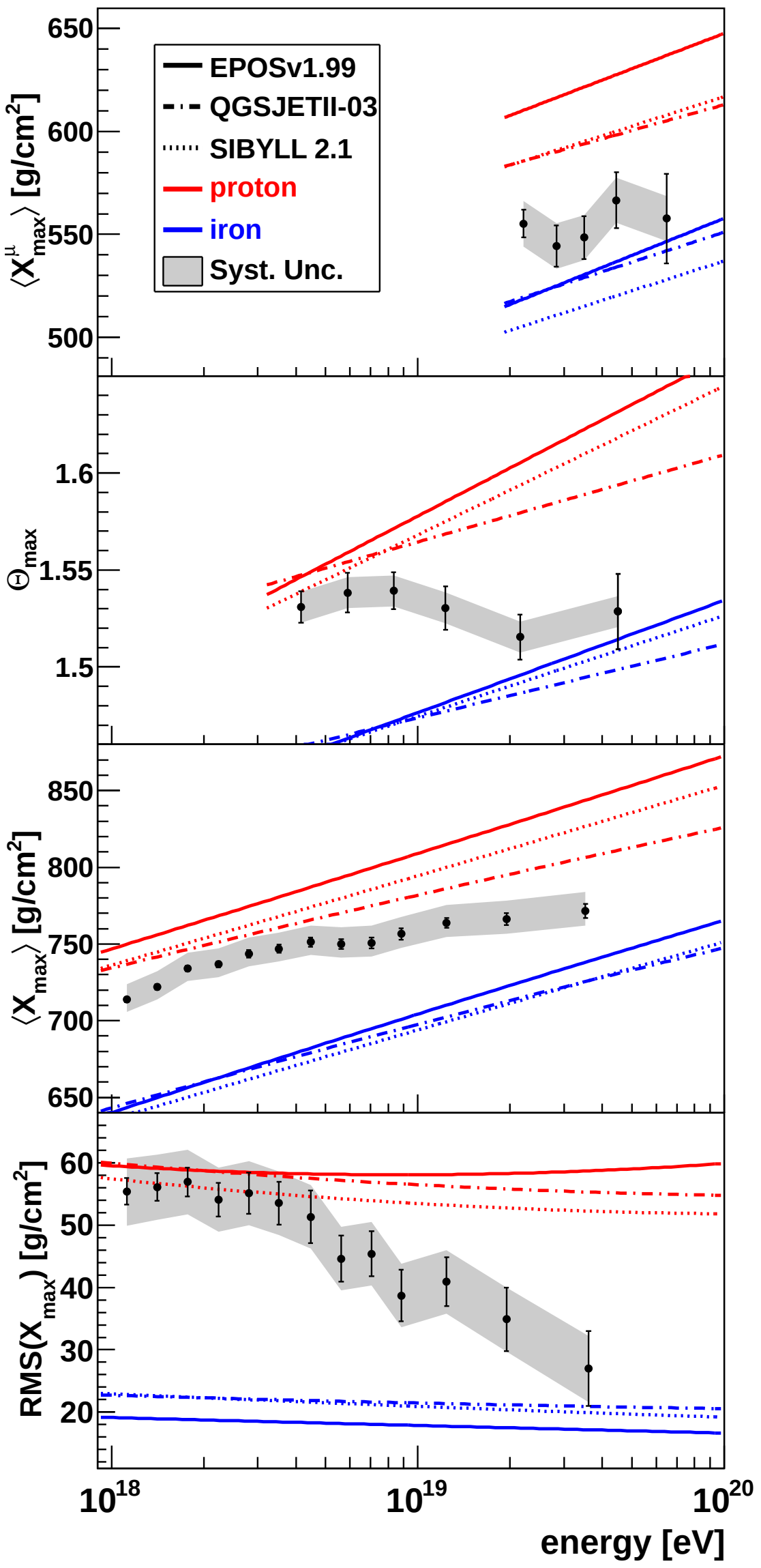

Figure 2: Results on shower evolution sensitive observables compared with models prediction. The error bars correspond to the statistical uncertainty. The systematic uncertainty is represented by the shaded bands.

Data collected with the surface detector of the Pierre Auger Observatory from January 2004 to December 2010 have been used for the $\Theta_{\rm max}$ analysis, with a total of 18581 events surviving the following cuts. Events are required to satisfy the trigger levels described in [9] and to be in the regime of full array efficiency for all primary species: $E > 3.16 \times 10^{18}$ eV and $\theta \leq 60°$. For selected events, detectors are used in the analysis if the signal size is above 10 VEM and not saturated and if they have core distances between 500 m and 2000 m. The measured values of $\Theta_{\rm max}$ obtained for 6 bins of energy above $3.16 \times 10^{18}$ eV are shown in Fig. 2. The systematic uncertainty in the measured values of $\Theta_{\rm max}$ has been evaluated taking into account its possible sources: reconstruction of the core of the shower, event selection and risetime vs core distance parameterisation and amounts to $\lesssim 10\%$ of the proton-iron separation predicted by the models. We note that muon numbers predicted by EAS simulations differ from those observed in data [2]. A preliminary study, using a normalization of 1.6 [2], indicates a possible change of about $\leq 5\%$ of the proton-iron difference.

As mentioned above, the shower observables $\Theta_{\rm max}$ and $X_{\rm max}$ are expected to be correlated as both are dependent upon the rate of shower development. The correlation between $\Theta_{\rm max}$ and $X_{\rm max}$ shown in Fig. 3 has been obtained with hybrid data using criteria similar to those adopted in [4]. In Fig. 3 the $\Theta_{\rm max}$ vs $X_{\rm max}$ correlations found with Monte Carlo data are also shown for proton and iron primaries, demonstrating that the correlation is independent of the primary mass.

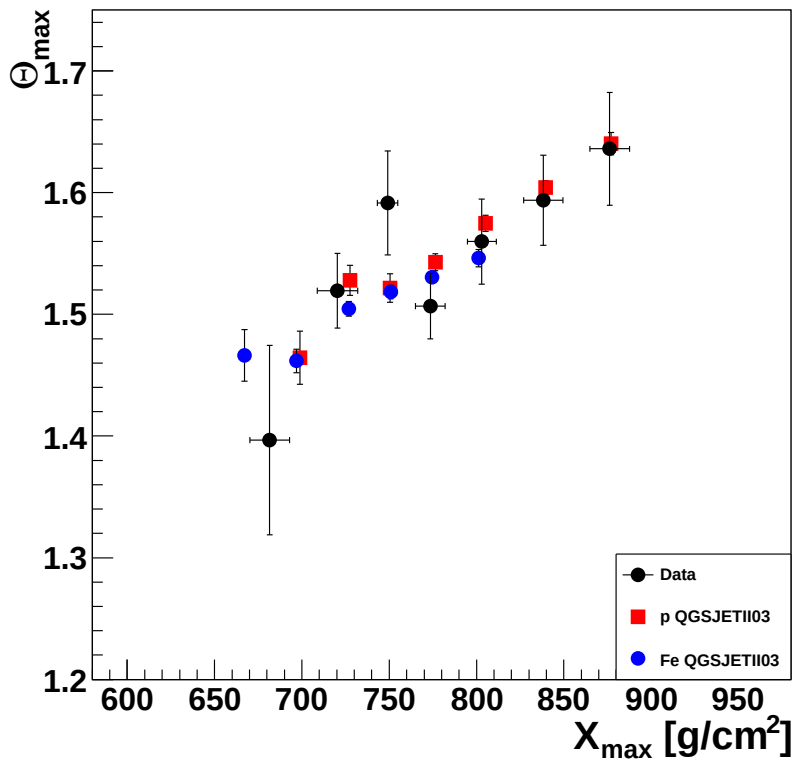

Figure 3: $\Theta_{\rm max}$ vs $X_{\rm max}$. Black dots correspond to data, while Monte Carlo results for proton(iron) primary are indicated by red(blue) squares(circles).

## 2.3 Depth Profile of Muon Production Points

Using the time information of the signals recorded by the SD it is also possible to obtain information about the longitudinal development of the hadronic component of extensive air showers in an indirect way. In [10] a method is presented to reconstruct the Muon Production Depth (MPD), i.e. the depth at which a given muon is produced measured parallel to the shower axis, using the FADC traces of detectors far from the core. The MPD technique allows us to convert the time distribution of the signal recorded by the SD detectors into muon production distances using an approximate relation between production distance, transverse distance and time delay with respect the shower front plane. From the MPDs an observable can be defined, $X^{\mu}_{\rm max}$, as the depth along the shower axis where the number of produced muons reaches a maximum. This new observable is a parameter sensitive to the longitudinal shower evolution which, as in the case of $\Theta_{\max}$, can be obtained with the information provided by the SD alone (see [11] for detailed explanation of the analysis). The method is currently restricted to inclined showers where muons dominate the signal at ground level (studies to extend the analysis to vertical showers are ongoing). Once the MPD is obtained for each event, the value of $X^{\mu}_{\max}$ is found by fitting a Gaisser-Hillas function to the depth profile. An example of the MPD profile and the result of the Gaisser-Hillas fit of a particular event with $E \approx 95$ EeV and zenith angle $\theta \approx 60°$ is shown in the right panel of Fig. 1.

The results of $\langle X^{\mu}_{\max}\rangle$ presented here are based on data collected between January 2004 and December 2010, with zenith angles between $55°$ and $65°$. The angular window was chosen as a trade-off between muon to EM ratio and the reconstruction uncertainty. The finite time resolution in the FADC traces produces an uncertainty on the reconstruction that decreases with the core distance and increases with the zenith angle. Thus, to keep these distortions low, only detectors far from the core (r $>$ 1800 m) can be used. This distance restriction imposes a severe limitation in the energy range where the method can by applied. Therefore only events with reconstructed energy larger than 20 EeV are used. After applying a set of reconstruction and quality cuts (see [11] for a complete description of the cuts), a total of 244 events are selected. The measured values of $\langle X^{\mu}_{\max}\rangle$ are presented in the upper panel of Fig. 2. The systematic uncertainty due to reconstruction bias, core position, rejection of the EM component and quality cuts amounts to $11\ \mathrm{g/cm^2}$, corresponding to about 14% of the proton-iron separation predicted by the models [11]. The predictions of $X^{\mu}_{\max}$ from different hadronic models (such as those shown in Fig. 2) would not be affected if a discrepancy between a model and data [2] is limited to the total number of muons. However, differences in the muon energy and spatial distribution would modify the predictions.

As for $\Theta_{\max}$, it is expected that the values of $X^{\mu}_{\max}$ will be correlated with $X_{\max}$. However there are insufficient events to make an experimental test such as that shown in Fig. 3. In Fig. 4 the results of model calculations are displayed using QGSJETII-03 as the hadronic model: the anticipated correlation is seen.

## 3 Conclusions

It is clear from Fig. 2 that if the models give a fair representation of the theoretical systematics of air shower modelling, then one might infer the primary composition from the data on the longitudinal air shower development presented here.

The evolution of $\langle X_{\max}\rangle$, $\Theta_{\max}$ and $\langle X^{\mu}_{\max}\rangle$ with energy is similar, despite the fact that the three analyses come from completely independent techniques that have different sources of systematic uncertainties. Concerning the RMS of $X_{\max}$, a variety of compositions can give rise to large values of the RMS, because the width of the $X_{\max}$ is influenced by both, the shower-to-shower fluctuations of individual components and their relative displacement in terms of $\langle X_{\max}\rangle$. However, within experimental uncertainties, the behaviour of $\langle X_{\max}\rangle$, $\Theta_{\max}$ and $\langle X^{\mu}_{\max}\rangle$ as shown in Fig. 2 is compatible with the energy evolution of RMS($X_{\max}$). In particular, at the highest energies all four analyses show consistently that our data resemble more the simulations of heavier primaries than pure protons.

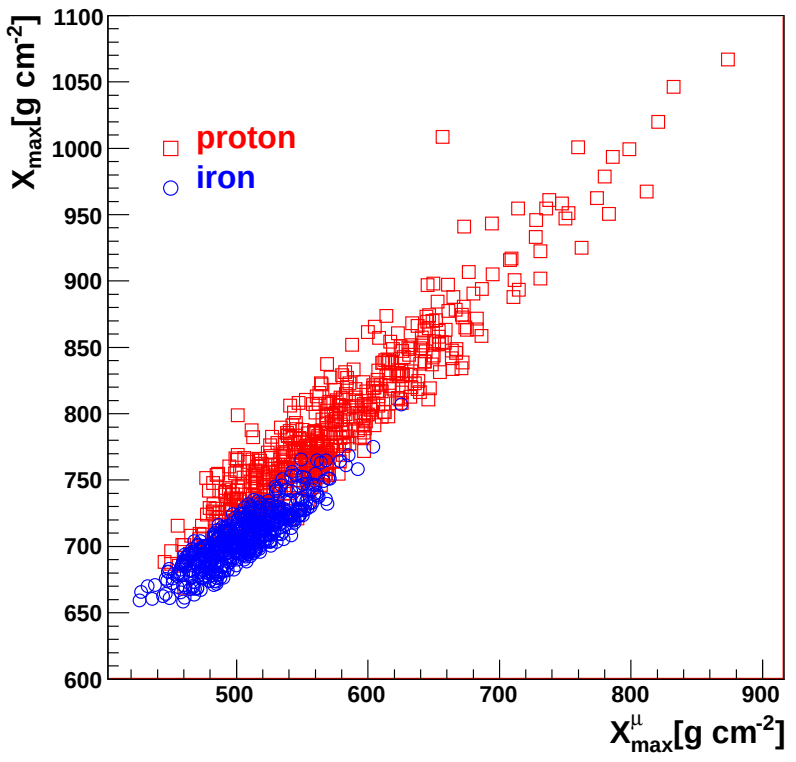

Figure 4: $X_{\max}$ vs $X^{\mu}_{\max}$ obtained for proton and iron simulated showers using QGSJETII-03 hadronic interaction model.

## References

[1] N. N. Kalmykov, S. S. Ostapchenko, and A. I. Pavlov, (1997) Nucl. Phys. Proc. Suppl. **52B**: 17-28; E.-J. Ahn *et al.*, (2009), Phys. Rev. **D80**: 094003; T. Pierog and K. Werner, (2008), Phys. Rev. Lett. **101**: 171101.

[2] J. Allen for the Pierre Auger Collaboration, paper 0703, these proceedings.

[3] The Pierre Auger Collaboration, (2004), Nucl. Instrum. Meth. **A523**: 50-95.

[4] The Pierre Auger Collaboration, (2010), Phys. Rev. Lett. **104**: 091101.

[5] P. Facal for the Pierre Auger Collaboration, paper 0725, these proceedings.

[6] H. Wahlberg for the Pierre Auger Collaboration, (2009), Proc. 31st ICRC, Łódź, Poland. arXiv:0906.2319v1[astro-ph].

[7] M.T. Dova for the Pierre Auger Collaboration, (2003), Proc. 28th ICRC, Tsukuba, Japan. 369-372.

[8] M.T. Dova *et al.*, (2009), Astropart. Phys. **31**: 312-319.

[9] The Pierre Auger Collaboration, (2010), Nucl. Instrum. Meth. **A613**: 29-39.

[10] L. Cazon, R.A. Vazquez and E. Zas, (2005), Astropart. Phys. **23**: 393-409.

[11] D. Garcia-Gamez for the Pierre Auger Collaboration, paper 0735, these proceedings.

[12] D. Heck *et al.*, (1998) Report **FZKA 6019** FZ Karlsruhe; S.J. Sciutto, (1999), arXiv:astro-ph/9911331.

# Estimate of the proton-air cross-section with the Pierre Auger Observatory

RALF ULRICH[1] FOR THE PIERRE AUGER COLLABORATION[2]
[1] *Karlsruhe Institute of Technology, Karlsruhe, Germany*
[2] *Observatorio Pierre Auger, Av. San Martin Norte 304, 5613 Malargüe, Argentina*
*(Full author list: http://www.auger.org/archive/authors_2011_05.html)*
*auger_spokespersons@fnal.gov*

**Abstract:** Using the tail of the distribution of the depth of shower maxima observed with the Pierre Auger Observatory, we derive an estimate of the proton-air cross-section for particle production at center-of-mass energies of $57$ TeV. Air showers observed with the fluorescence detector and at least one station of the surface detector array are analysed in the energy range from $10^{18}$ to $10^{18.5}$ eV. Systematic uncertainties in the cross-section estimate arising from the limited knowledge of the primary mass composition, the need to use shower simulations and the selection of showers are studied in detail.

**Keywords:** Proton-air, Cross-section, Pierre Auger Observatory

## 1 Introduction

One of the biggest challenges towards a better understanding of the nature of ultra-high energy cosmic rays is to improve the modeling of hadronic interaction in air showers. Currently, none of the models is able to consistently describe cosmic ray data, which most importantly prevents a precise determination of the primary cosmic ray mass composition.

Studies to exploit the sensitivity of cosmic ray data to the characteristics of hadronic interactions at energies beyond state-of-the-art accelerator technology began over $50$ years ago. While first measurements were based on the direct observation of cosmic ray particles [1], the rapidly shifting focus towards higher energies required the use of extensive air shower observations [2, 3]. The property of interactions most directly linked to the development of extensive air showers is the cross-section for the production of hadronic particles (e.g. [4, 5]).

We present the first analysis of the proton-air cross-section based on hybrid data from the Pierre Auger Observatory. For this purpose we analyse the shape of the distribution of the largest values of the depth of shower maxima, $X_{\rm max}$, the position at which air showers deposit the maximum energy per unit of mass of atmosphere traversed. This *tail* of the $X_{\rm max}$-distribution is very sensitive to the proton-air cross-section, a technique first exploited in the pioneering work of Fly's Eye [3]. To obtain accurate measurements of $X_{\rm max}$, the timing data from the fluorescence telescopes is combined with that from the surface detector array for a precise reconstruction of the geometry of events.

An over-riding concern of the analysis has been the assignment of realistic systematic uncertainties to the result. We recognise and identify the unknown mass composition of cosmic rays as *the* major source of systematic uncertainty for the proton-air cross-section analysis and we evaluate its impact on the final result. The analysis is optimised to minimise the impact of contamination by the presence of particles other than protons in the primary beam.

## 2 Analysis approach

The method used to estimate the proton-air cross-section is the comparison of an appropriate air shower observable with Monte Carlo predictions. A disagreement between data and predictions is then attributed to a modified value of the proton-air cross-section. The present analysis is a two-step process.

Firstly, we measure an air shower observable with high sensitivity to the cross-section. Secondly, we convert this measurement into an estimate of the proton-air cross-section for particle production, $\sigma_{\rm p-air}$, in the energy interval $10^{18}$ to $10^{18.5}$ eV. The selection of this energy range has the following advantages and features:

**Statistics:** A large number of events are recorded.

**Composition:** The shape of the $X_{\rm max}$-distribution is compatible with there being a substantial fraction of protons in the cosmic ray beam. The situation is less clear at higher energies.

**Energy:** The average center-of-mass energy for a cosmic ray proton interacting with a nucleon in the atmosphere is 57 TeV, significantly above what will be reached at the LHC.

As the primary observable we define $\Lambda_f$ via the exponential shape $\mathrm{d}N/\mathrm{d}X_{\max} \propto \exp(-X_{\max}/\Lambda_f)$ of the $X_{\max}$-distribution of the fraction $f$ of the most deeply penetrating air showers. Considering only these events enhances the contribution of protons in the sample as the average depth at which showers maximise is higher in the atmosphere for non-proton primaries.

The choice of the fraction $f$ is a crucial part of the definition of the observable $\Lambda_f$. While a small value of $f$ will enhances the proton fraction, since protons penetrate most deeply of all primary nuclei, it also reduces the number of events for the analysis. By varying $f$ we investigate how much the bias due to non-proton induced showers can be reduced without statistical uncertainties being dominant. Following these studies we have chosen $f = 20\,\%$ so that for helium-fractions up to 15 % biases induced by helium are kept below the level of the statistical resolution. At the same time this choice suppresses elements heavier than helium very efficiently.

## 3 The Measurement of $\Lambda_f$

We use events collected between December 2004 and September 2010. The atmospheric and event-quality cuts applied are identical to those used for the analysis of $\langle X_{\max}\rangle$ and RMS($X_{\max}$) [6,8]. This results in $11\,628$ high quality events between $10^{18}$ and $10^{18.5}$ eV.

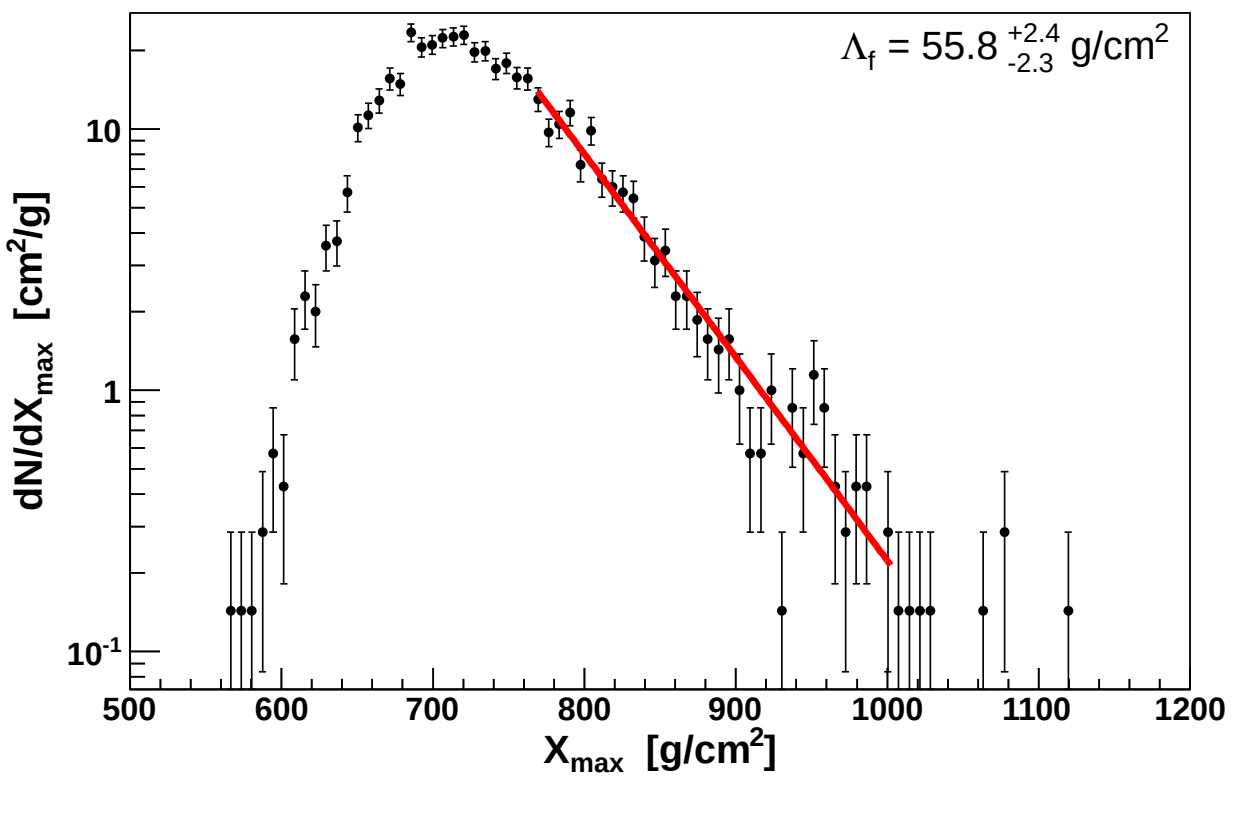

Figure 1: Unbinned likelihood fit of $\Lambda_f$ to the tail of the $X_{\max}$ distribution.

The $X_{\max}$ distribution of the data is affected by the known geometrical acceptance of the fluorescence telescopes as well as by detection limitations related to atmospheric light transmission. The impact of the telescope acceptance on the $X_{\max}$ distribution is well understood and can be studied by using data (see [8]) and with detailed Monte Carlo simulations of the shower detection process. In the following we use the strategy developed for the measurement of the $\langle X_{\max}\rangle$ and RMS($X_{\max}$) [6,8] to extract a data sample that has an unbiased $X_{\max}$ distribution.

In the first step we derive the range of values of $X_{\max}$ that corresponds to the deepest $f = 20\,\%$ of the measured showers. We select only event geometries that allow, for each shower, the complete observation of the slant depth range from $550$ to $1004\,\mathrm{g/cm^2}$, which corresponds to 99.8 % of the observed $X_{\max}$-distribution. These fiducial volume cuts reduce the data sample to $1\,635$ events, providing a good estimate of the unbiased $X_{\max}$-distribution. This distribution is then used to find the range of values of $X_{\max}$ that contains the $20\,\%$ deepest showers, which is identified to extend from 768 to 1004 g/cm$^2$. Due to the limited statistics involved in this range estimation, there is a $\pm 1.5\,\mathrm{g/cm^2}$ uncertainty on the definition of the range of the tail, which will be included in the estimation of the systematic uncertainties.

In the second step we select those events from the original data sample of $11\,628$ high quality events that allow the complete observation of values of $X_{\max}$ from 768 to 1004 g/cm$^2$, corresponding to the $20\,\%$-tail of the unbiased distribution. This is a more relaxed fiducial volume cut since we are not requiring that a shower track can be observed at depths higher in the atmosphere than 768 g/cm$^2$, which maximises the event statistics and still guarantees an unbiased $X_{\max}$ distribution in the range of interest. In total there are $3\,082$ showers passing the fiducial volume cuts, of which 783 events have their $X_{\max}$ in the selected range and thus directly contribute to the measurement of $\Lambda_f$. The average energy of these events is $10^{18.24}$ eV, corresponding to a center-of-mass energy of $\sqrt{s} = 57$ TeV in proton-proton collisions.

In Fig. 1 we show the data and the result of an unbinned maximum likelihood fit of an exponential function over the range 768 to $1004\,\mathrm{g/cm^2}$. This yields

$$\Lambda_f = (55.8 \pm 2.3_{\rm stat} \pm 0.6_{\rm syst})\ \mathrm{g/cm^2}\ . \qquad (1)$$

The systematic uncertainty arises from the precision with which the range of depths that are used can be defined.

Values of $\Lambda_f$ have been calculated for modified event selections and for different ranges of atmospheric depths. It is found that the changes in $\Lambda_r$ lie within the statistical uncertainties. The re-analysis of sub-samples selected according to zenith-angle, shower-telescope distance and energy produces variations of the value of $\Lambda_f$ consistent with statistical fluctuations. We conclude that the systematic uncertainties related to the measurement are below $5\,\%$.

## 4 Determination of the cross-section

We must resort to Monte Carlo simulations to derive an estimate of the proton-air cross-section from the measurement of $\Lambda_f$. These have been made using the same energy distribution as in the data, and the events from the simulations have been analysed with the identical procedures used for the data.

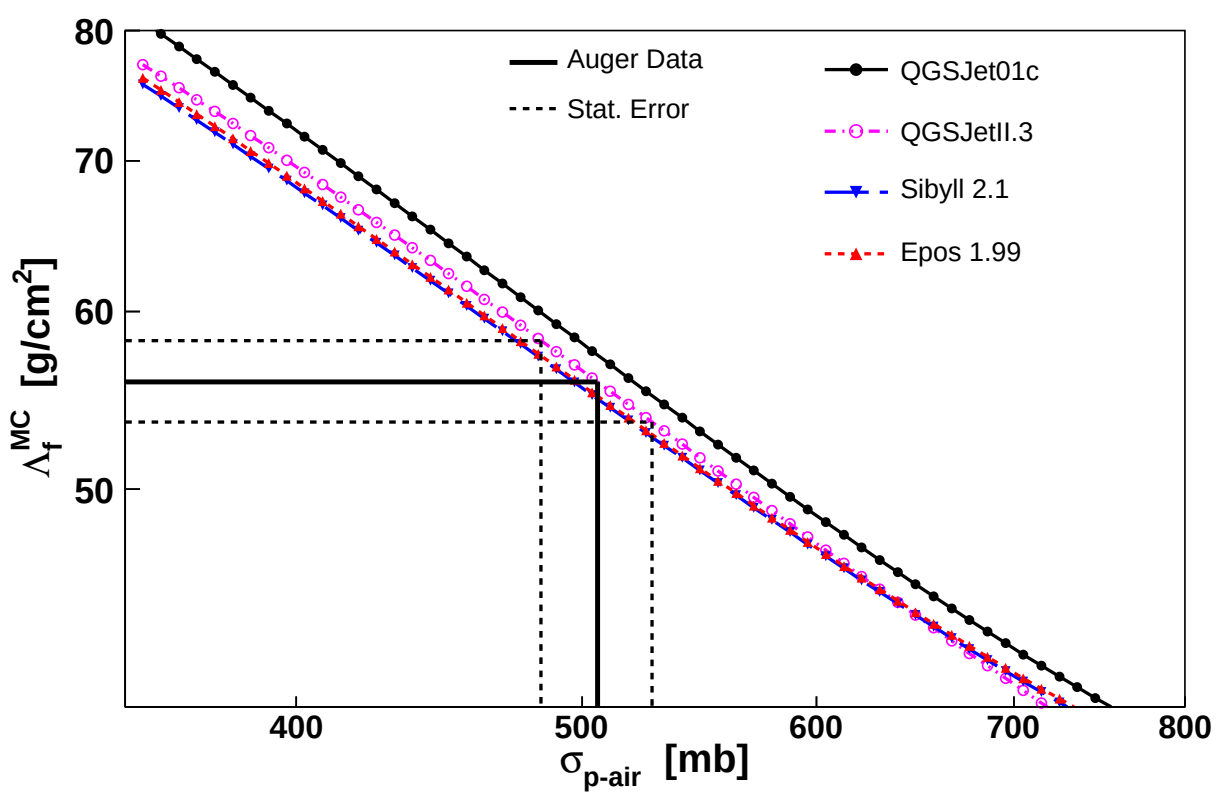

Figure 2: Relation between $\Lambda_f^{\mathrm{MC}}$ and $\sigma_{\mathrm{p-air}}$. As example we show the conversion of the measurement $\Lambda_f^{\mathrm{MC}} = \Lambda_f$ with the QGSJetII model.

Table 1: Cross-sections derived from the measured $\Lambda_f$ using different interaction models. The given uncertainties are statistical only. The rescaling factor, $m(E,\,f_{19})$, is a measure of how much the original cross-section of the model have to be changed.

| Model | Rescaling factor at $10^{18.24}$ eV | $\sigma_{\mathrm{p-air}}$/mb |
|---|---|---|
| QGSJet01 | $1.04 \pm 0.04$ | $524 \pm 23$ |
| QGSJetII.3 | $0.95 \pm 0.04$ | $503 \pm 22$ |
| SIBYLL 2.1 | $0.88 \pm 0.04$ | $497 \pm 23$ |
| EPOS 1.99 | $0.96 \pm 0.04$ | $498 \pm 22$ |

In general, the Monte Carlo values of $\Lambda_f^{\mathrm{MC}}$ do not agree with the measurement. It is known from previous work that the values of $\Lambda_f^{\mathrm{MC}}$ derived from simulations are directly linked to the hadronic cross-sections used in the simulations. Accordingly we can explore the effect of changing cross-sections in an empirical manner by multiplying the cross-sections that are input to the simulations by an energy-dependent factor [7]

$$m(E,\,f_{19}) = 1 + (f_{19} - 1)\,\frac{\lg\left(E/10^{15}\,\mathrm{eV}\right)}{\lg\left(10^{19}\,\mathrm{eV}/10^{15}\,\mathrm{eV}\right)}, \quad (2)$$

where $E$ denotes the shower energy and $f_{19}$ is the factor by which the cross-section is rescaled at $10^{19}$ eV. The rescaling factor is unity below $10^{15}$ eV reflecting the fact that measurements of the cross-section at the Tevatron were used for tuning the interaction models. This technique of modifying the original cross-sections predictions during the Monte Carlo simulation process assures a smooth transition from accelerator data up to the energies of our analysis. For each hadronic interaction model, the value of $f_{19}$ is obtained that reproduces the measured value of $\Lambda_f$. The cross-section is then deduced by multiplying the factor Eq. (2) to the original model cross-section.

In Fig. 2 we show the conversion curves for simulations based on the four most commonly used high-energy hadronic interaction models for air shower simulations (Sibyll2.1 [9], QGSJet01 [10], QGSJetII.3 [11] and EPOS1.99 [12]).

The need to use Monte Carlo calculations introduces model-dependence to this section of the analysis. It is known that other features of hadronic interactions, such as the multiplicity and elasticity, have an impact on air shower development [4, 5]. We use the very different multiparticle production characteristics of the four models to sample the systematic effect induced by these features.

The proton-air cross-sections for particle production derived are given in Table 1. Only SIBYLL needs to be modified with a rescaling factor significantly different from unity to describe the tail of the measured $X_{\max}$ distribution.

The systematic uncertainty of 22 % [13] in the absolute value of the energy scale leads to systematic uncertainties of 7 mb in the cross-section and 6 TeV in the center-of-mass energy.

Furthermore, the simulations needed to obtain $\sigma_{\mathrm{p-air}}$ from the measured $\Lambda_f$ as shown in Fig. 2 depend on additional parameters. By varying for example the energy distribution, energy and $X_{\max}$ resolution of the simulated events, we find that related systematic effects are below 7 mb.

The average depth of $X_{\max}$ of showers produced by photons in the primary beam at the energies of interest lies about $50\,\mathrm{g/cm^2}$ deeper in the atmosphere than for protons. The presence of photons would bias the measurement. However, observational limits on the fraction of photons are $< 0.5$ % [14, 15] and the corresponding underestimation of the cross-section is less than 10 mb.

With the present limitations of air shower observations, it is impossible to distinguish showers that are produced by helium nuclei from those created by protons. Accordingly, lack of knowledge of the helium fraction leads to a significant systematic uncertainty. From simulations we find that $\sigma_{\mathrm{p-air}}$ is overestimated by 10, 20, 30, 40 and 50 mb for percentages of helium of 7.5, 20, 25 32.5 and 35% respectively. We find that CNO-group nuclei introduce no bias for fractions up to $\sim 50$ %, thus we assign no systematics on the cross-section for it.

In Table 2, where the systematic uncertainties are summarised, we quote results for 10, 25 and 50 % of helium.

Table 2: Summary of the systematic uncertainties.

| Description | Impact on $\sigma_{\mathrm{p-air}}$ |
|---|---|
| $\Lambda_r$ systematics | $\pm 6$ mb |
| Hadronic interaction models | $^{+16}_{-9}$ mb |
| Energy scale | $\pm 7$ mb |
| Simulations and parameterisations | $\pm 7$ mb |
| Photons, <0.5 % | <+10 mb |
| Helium, 10 % | -12 mb |
| Helium, 25 % | -30 mb |
| Helium, 50 % | -80 mb |
| Total (w/o composition) | -15 mb, +20 mb |

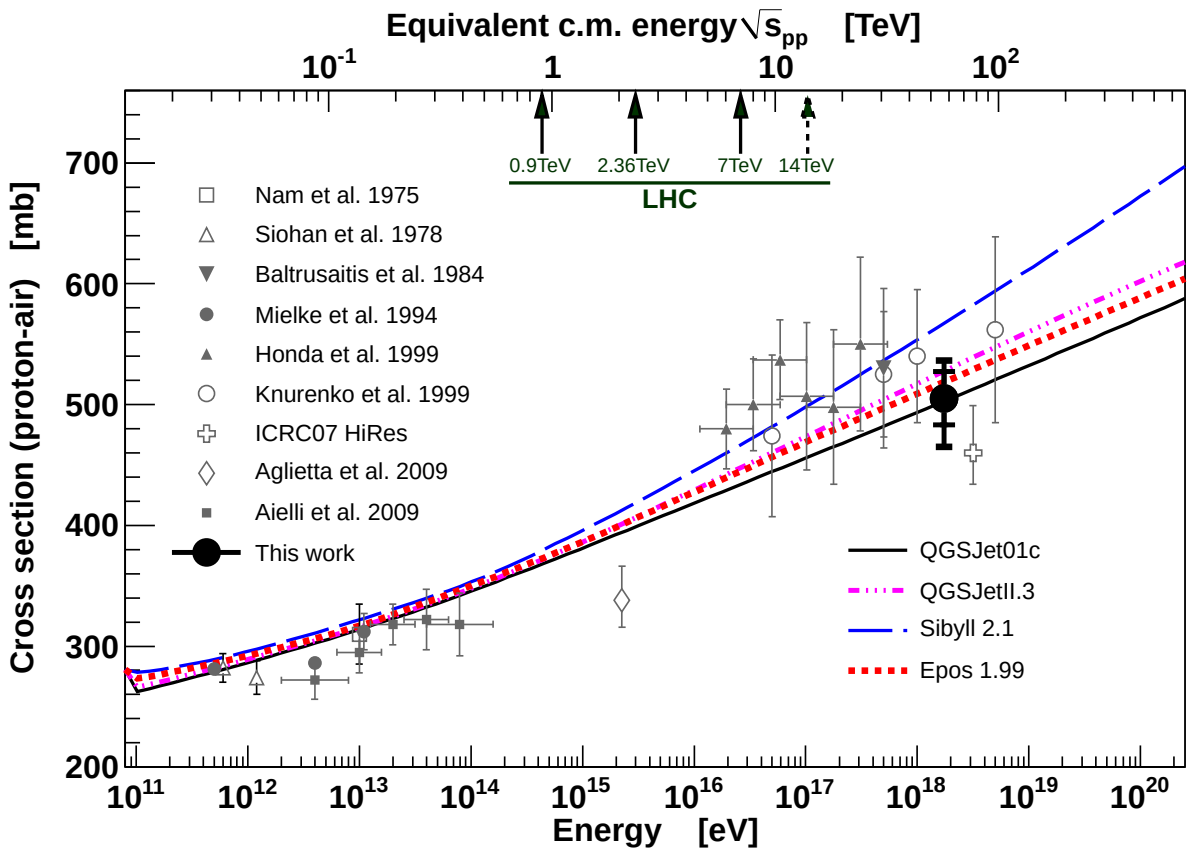

Figure 3: Resulting $\sigma_{\mathrm{p-air}}$ compared to other measurements [1–3] and model predictions. The inner error bars are statistical only, while the outer include all systematic uncertainties for a helium fraction of 25 % and 10 mb photon systematics.

It is interesting to note that the model-dependence is moderate and does not dominate the measurement.

We summarise our results by averaging the four values of the cross-section (Table 1) to give

$$\sigma_{\mathrm{p-air}} = \left(505\ \pm 22_{\mathrm{stat}}\ \left({}^{+20}_{-15}\right)_{\mathrm{syst}}\right)\ \mathrm{mb}$$

at a center-of-mass energy of $57 \pm 6$ TeV. The helium-induced systematics is -12, -30 and -80 mb for 10, 25 and 50 % of helium, respectively and the photon-induced bias $<$+10 mb. In Fig. 3 we compare this result with model predictions and other measurements derived from cosmic ray data.

## 5 Discussion

We have developed a method to determine the cross-section for the production of particles in proton-air collisions from data of the Pierre Auger Observatory. We have studied in detail the effect of the primary cosmic ray mass composition, hadronic interaction models, simulation settings and telescope fiducial volume limits on the final result. The fundamental assumption for the analysis is that the light cosmic ray mass component in the selected data set is dominated by proton primaries. The systematic uncertainties arising from the lack of knowledge of the helium and photon components are potentially the largest source of systematic uncertainty. However, for helium fractions up to 25 % the induced bias remains small. One could also argue that only a specific amount of helium is allowed in the data since otherwise the hadronic cross-sections at ultra-high energies would become very small and at some point inconsistent with the extrapolation of accelerator data to $\sqrt{s} = 57$ TeV.

Our result favours a moderately slow rise of the cross-section towards higher energies. This has implications for expectations at the LHC. First analyses at the LHC also indicate slightly smaller hadronic cross-sections than expected within many models [16].

We plan to convert the derived $\sigma_{\mathrm{p-air}}$ measurement into the more fundamental cross-section of proton-proton collisions using the Glauber framework [17]. Thus, a direct comparison to accelerator measurements will be possible.

## References

[1] G. Yodh *et al.*, Phys. Rev. Lett., 1972, **28**: 1005. R. Nam *et al.*, Proc. of 14th Int. Cosmic Ray Conf., Munich, 1975, **7**: 2258. F. Siohan *et al.*, J. Phys., 1978, **G4**: 1169. H. Mielke *et al.*, J. Phys., 1994, **G20**: 637.

[2] M. Honda *et al.*, Phys. Rev. Lett., 1993, **70**: 525. M. Aglietta *et al.*, EAS-TOP Collaboration, Phys. Rev., 2009, **D79**: 032004. T. Hara *et al.*, Phys. Rev. Lett., 1983, **50**: 2058. G. Aielli *et al.*, ARGO Collaboration, Phys. Rev., 2009, **D80**: 092004. S. Knurenko *et al.*, Proc. of 26th Int. Cosmic Ray Conf., Salt Lake City, 1999, **1**: 372.

[3] R. Ellsworth *et al.*, Phys. Rev., 1982, **D26**: 336. R. Baltrusaitis *et al.*, Phys. Rev. Lett., 1984, **52**: 1380.

[4] J. Matthews, Astropart. Phys., 2005, **22**: 387.

[5] R. Ulrich *et al.*, Phys. Rev., 2011, **D83**: 054026.

[6] P. Facal for the Pierre Auger Collaboration, paper 0725, these proceedings.

[7] R. Ulrich *et al.*, New J. Phys., 2009, **11**: 065018.

[8] J. Abraham *et al.*, Pierre Auger Collaboration, Phys. Rev. Lett., 2010, **104**: 091101.

[9] E. Ahn *et al.*, Phys. Rev., 2009, **D80**: 094003.

[10] N. Kalmykov and S. Ostapchenko, Phys. Atom. Nucl., 1993, **56**: 346.

[11] S. Ostapchenko, Phys. Rev., 2006, **D74**: 014026.

[12] K. Werner, Phys. Rev., 2006, **C74**: 044902.

[13] R. Pesce for the Pierre Auger Collaboration, paper 1160, these proceedings.

[14] A. Glushkov *et al.*, Phys. Rev., 2010, **D82**: 041101.

[15] M. Settimo for the Pierre Auger Collaboration, paper 0393, these proceedings.

[16] G. Aad *et al.*, ATLAS Collaboration, 2011, arXiv:1104.0326 [hep-ex]. CMS Collaboration, presentation at DIS workshop, Brookhaven, 2011.

[17] R. Glauber, Phys. Rev., 1955, **100**: 242. R. Glauber and G. Matthiae, Nucl. Phys., 1970, **B21**:135.

# Interpretation of the signals produced by showers from cosmic rays of $10^{19}$ eV observed in the surface detectors of the Pierre Auger Observatory

JEFF ALLEN[1] FOR THE PIERRE AUGER COLLABORATION[2]
[1]*New York University, 4 Washington Place, New York, NY, 10003, USA*
[2]*Observatorio Pierre Auger, Av. San Martin Norte 304, 5613 Malargüe, Argentina*
*(Full author list: http://www.auger.org/archive/authors_2011_05.html)*
*auger_spokespersons@fnal.gov*

**Abstract:** Muons in extensive air showers are messengers of the hadronic-shower core whose simulation is subject to large theoretical uncertainties due to our limited knowledge of multi-particle production in hadronic interactions. Different methods of deriving the fraction of the signal observed in the surface detectors coming from either the muonic or electromagnetic shower components are used to compare the data from the Pierre Auger Observatory with predictions of Monte Carlo simulations. The observations are quantified relative to the predictions obtained with QGSJET II and FLUKA as interaction models. The predicted number of muons at 1000 m from the shower axis is lower than that found in data, and the energy that would have to be assigned to the surface detector signal, based on shower simulations, is systematically higher than that derived from fluorescence observations.

**Keywords:** muons, hadronic interactions, ultra high energy extensive air showers, muon deficit, simulations

## 1 Introduction

The Pierre Auger Observatory is a powerful detector for studying extensive air showers at very high energy. The combination of the fluorescence detector (FD) and surface detector array (SD) of the Observatory allows the simultaneous measurement of several observables of showers, providing opportunities to cross-check our current understanding of the physics of air showers. Many features of air showers depend directly on the characteristics of hadronic interactions which are unknown at very high energy and in phase space regions not covered in accelerator experiments. For example, recent work has quantified the sensitivity of the number of muons in ultra-high-energy air showers to several properties of hadronic interactions, including the multiplicity, the charge ratio (the fraction of secondary pions which are neutral), and the baryon anti-baryon pair production [1, 2]. Using models of hadronic interactions that do not provide a good description of shower data might lead to incorrect conclusions about the mass and the energy assignment being drawn from measurements.

In this work the data of the Pierre Auger Observatory is compared to showers simulated using the interaction model QGSJET II.03 [3], which has become a standard reference model for air-shower experiments. Updates are provided to several methods presented previously [4], and a new method is introduced. In Sec. 2, the data from the surface and fluorescence detectors is compared simultaneously, on an event-by-event basis, to the results of simulations. In Sec. 3, the time structure of the particle signals in the surface detectors and a universal property of air showers are used to estimate the number of muons in the data. Finally, in Sec. 4, the ground signals of simulated events are matched to those measured by rescaling the number of muons arising from hadronic processes and changing the energy assignment in simulated showers.

## 2 Study of Individual Hybrid Events

At the Auger Observatory, thousands of showers have been recorded for which reconstruction has been possible using both the FD and SD. These hybrid events have been used to construct a library of simulated air-shower events where the longitudinal profile (LP) of each simulated event matches a measured LP. The measured LP constrains the natural shower-to-shower fluctuations of the distribution of particles at ground. This allows the ground signals of simulated events to be compared to the ground signals of measured events on an event-by-event basis.

Hybrid events were selected using the criteria adopted for the energy calibration of the SD [5] in the energy range $18.8 < \log(E) < 19.2$ recorded between 1 January 2004 and 31 December 2008. 227 events passed all cuts. Air showers were simulated using SENECA [6] with QGSJET II and FLUKA [7] as the high- and low-energy event generators. For every hybrid event, three proton- and three iron-initiated showers were selected from a set of 200 simulated showers for each primary type. The energy and arrival direction of the showers matches the measured event, and the LPs of the selected showers have the lowest $\chi^2$ compared to the measured LP. The measured LP and two selected LPs of an example event are shown in the top panel of Fig. 1.

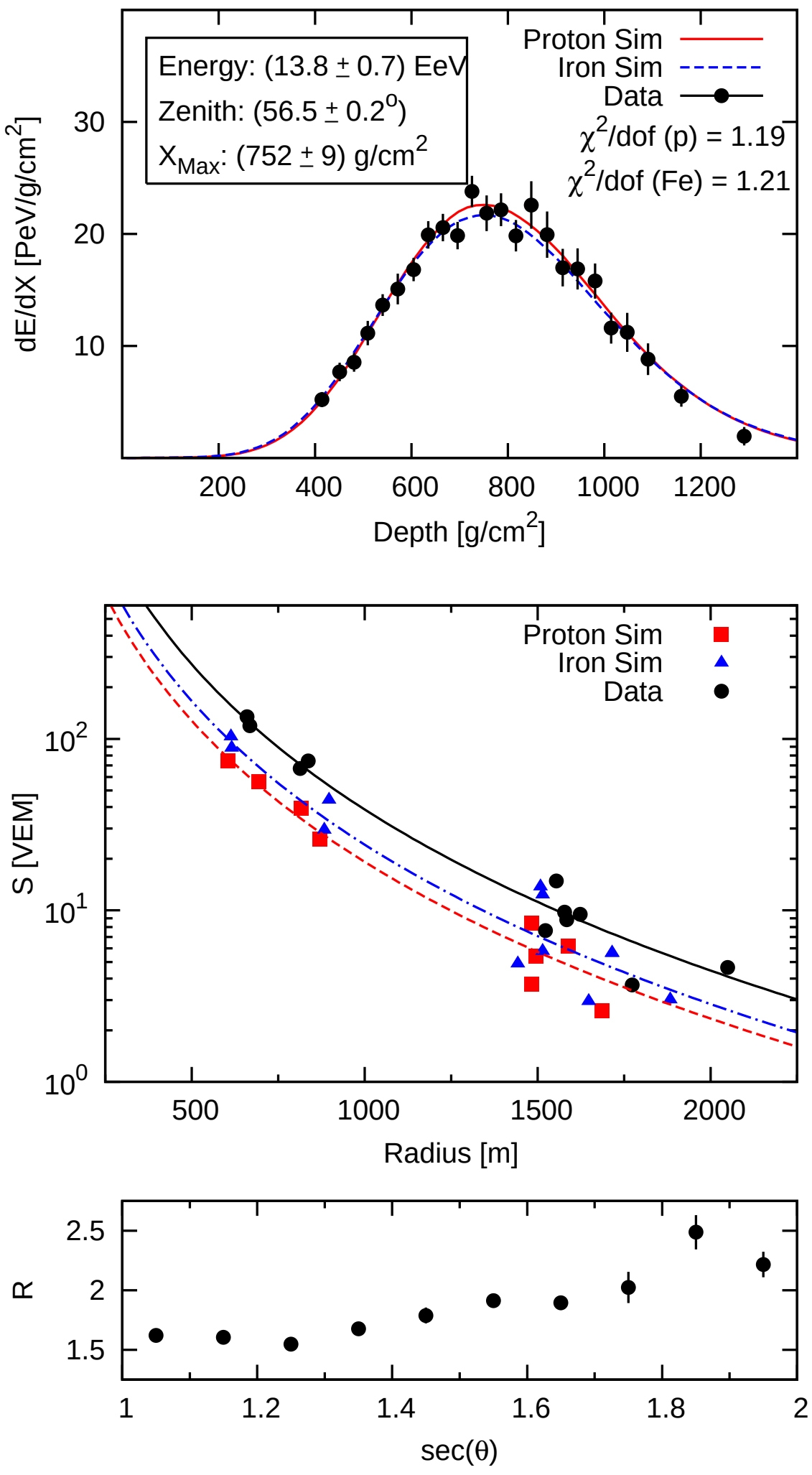

Figure 1: *Top panel:* A longitudinal profile measured for a hybrid event and matching simulations of two showers with proton and iron primaries. *Middle panel:* A lateral distribution function determined for the same hybrid event as in the top panel and that of the two simulated events. *Bottom panel:* $R$, defined as $\frac{\mathrm{S}(1000)_{\mathrm{Data}}}{\mathrm{S}(1000)_{\mathrm{Sim}}}$, averaged over the hybrid events as a function of sec$\theta$.

The detector response for the selected showers was simulated using the Auger $\overline{\text{Off}}\underline{\text{line}}$ software package [8, 9]. The lateral distribution function of an observed event and that of two simulated events are shown in the middle panel of Fig. 1. For each of the 227 events, the ground signal at 1000 m from the shower axis, S (1000), is smaller for the simulated events than that measured. The ratio of the measured S (1000) to that predicted in simulations of showers with proton primaries, $\frac{\mathrm{S}(1000)_{\mathrm{Data}}}{\mathrm{S}(1000)_{\mathrm{Sim}}}$, is 1.5 for vertical showers and grows to around 2 for inclined events; see the bottom panel of Fig. 1. The ground signal of more-inclined events is muon-dominated. Therefore, the increase of the discrepancy with zenith angle suggests that there is a deficit of muons in the simulated showers compared to the data. The discrepancy exists for simulations of showers with iron primaries as well, which means that the ground signal cannot be explained only through composition.

## 3 Estimate of the Muonic Signal in Data

### 3.1 A multivariate muon counter

In this section, the number of muons at 1000 m from the shower axis is reconstructed. This was accomplished by first estimating the number of muons in the surface detectors using the characteristic signals created by muons in the PMT FADC traces and then reconstructing the muonic lateral distribution function (LDF) of SD events.

In the first stage, the number of muons in individual surface detectors is estimated. As in the *jump method* [4], the total signal from discrete jumps

$$J = \sum_{\text{FADC bin } i} \underbrace{(x_{i+1} - x_i)}_{\text{jump}} \mathbb{I}\left\{x_{i+1} - x_i > 0.1\right\} \qquad (1)$$

was extracted from each FADC signal, where $x_i$ is the signal measured in the $i$th bin in Vertical Equivalent Muon (VEM) units, and the indicator function $\mathbb{I}\{y\}$ is 1 if its argument $y$ is true and 0 otherwise. The estimator $J$ is correlated with the number of muons in the detector, but it has an RMS of approximately 40%. To improve the precision, a multivariate model was used to predict the ratio $\eta = (N_\mu + 1)/(J + 1)$. 172 observables that are plausibly correlated to muon content, such as the number of jumps and the rise-time, were extracted from each FADC signal. Principal Component Analysis was then applied to determine 19 linear combinations of the observables which best capture the variance of the original FADC signals. Using these 19 linear combinations, an artificial neural network (ANN) [10] was trained to predict $\eta$ and its uncertainty. The output of the ANN was compiled into a probability table $P_{\text{ANN}} = P(N_\mu = N \,|\, \text{FADC signal})$. The RMS of this estimator is about 25%, and biases are also reduced compared to the estimator $J$.

In the second stage of the reconstruction, a LDF

$$\overline{N}(r, \nu, \beta, \gamma) = \exp\left(\nu + \beta \log\frac{r}{1000\,\text{m}} + \gamma \log\left(\frac{r}{1000\,\text{m}}\right)^2\right) \qquad (2)$$

is fit to the estimated number of muons in the detectors for each event, where $r$ is the distance of the detector from the shower axis and $\nu$, $\beta$, and $\gamma$ are fit parameters. The number of muons in each surface detector varies from the LDF according to the estimate $P_{\text{ANN}}$ and Poisson fluctuations. The fit parameters, $\nu$, $\beta$, and $\gamma$, have means which depend on the primary energy and zenith angle as well as variances arising from shower-to-shower fluctuations. Gaussian prior distributions with energy- and zenith-dependent means were defined for the three fit parameters. All the parameters were estimated using an empirical Bayesian approach: three iterations were performed between (i) finding the maximum *a posteriori* estimate $\widehat{\nu}_i$, $\widehat{\beta}_i$, and $\widehat{\gamma}_i$ for each shower $i$ given the fixed priors, and (ii) re-estimating the priors given the fixed parameter estimates $\widehat{\nu}_i$, $\widehat{\beta}_i$, and $\widehat{\gamma}_i$.

The value of the muonic LDF at 1000 m, $\exp(\widehat{\nu})$, is highly correlated with $N_\mu(1000)$, the number of muons in surface detectors 1000 m from the shower axis. The RMS of $\exp(\widehat{\nu})$ in showers simulated using QGSJET II is 12% and 5% for proton and iron primaries. To correct several biases that depend on the energy and zenith angle of the showers, a quadratic function $f\big(\exp(\widehat{\nu}), \widehat{\theta}\big)$ was tuned on a library of showers simulated using QGSJET II with simulated detector response. The final estimator $\widehat{N}_\mu(1000) = f\big(\exp(\widehat{\nu}), \widehat{\theta}\big)$ has a systematic uncertainty below 50° of 6% from uncertainty in the composition and $^{+10\%}_{-0\%}$ from uncertainty in the hadronic models, determined by reconstructing showers simulated using EPOS 1.6. The total systematic uncertainty decreases with the zenith angle: at $\theta = 55°$ it is $^{+9\%}_{-3\%}$.

### 3.2 Universality of $S_\mu/S_{\rm em}$ behavior on $X^{\rm v}_{\rm max}$

The ratio of the muonic signal to the electromagnetic (EM) signal, $S_\mu/S_{\rm em}$, at 1000 m from the shower axis exhibits an empirical universal property for all showers at a fixed vertical depth of shower maximum, $X^{\rm v}_{\rm max}$ [11]. $S_\mu/S_{\rm em}$ is independent of the primary particle type, primary energy, and incident zenith angle. The dependence of $S_\mu/S_{\rm em}$ on $X^{\rm v}_{\rm max}$ can be described by a simple parameterization which leads to the following expression for the muonic signal in showers with zenith angle between 45° and 65°

$$S_\mu^{\rm fit} = \frac{{\rm S}\,(1000)}{1 + \cos^\alpha(\theta)/\left((X^{\rm v}_{\rm max}/A)^{1/b} - a\right)}\,, \qquad (3)$$

where S (1000) is as defined above, $\theta$ is the zenith angle, $\alpha = 1.2$, and $A$, $a$ and $b$ are fit parameters [12]. The estimation of the muonic signal in data is complicated by the dependence of the fit parameters $A$, $a$, and $b$ in Eq. (3) on the choice of hadronic interaction model. This dependence gives rise to a systematic uncertainty in the measurement of the muonic signal in data which is difficult to quantify due to the uncertainty in properties of hadronic interactions. This problem can be overcome with an additional phenomenological consideration: for showers with zenith angles above 45°, the fraction of the EM signal coming from the decay and interactions of muons rapidly increases. As shown in [13], different models of hadronic interactions are in agreement on the $S_\mu/S_{\rm em}$ ratio for showers with zenith angle above 45°, since $S_\mu/S_{\rm em}$ increasingly reflects the equilibrium between muons and their EM halo.

The fit in Eq. (3) provides an unbiased estimate of both the muonic and EM signals. The RMS of the muonic signal in showers simulated using CORSIKA [14] is less than 5% and 3% for proton and iron primaries [12]. The systematic uncertainty of $S_\mu$ from uncertainty in the hadronic models is estimated to be 6%, determined by the application of the parameterization of Eq. (3) using QGSJET II on showers simulated using EPOS 1.99. The systematic uncertainty of $S_\mu$ from event reconstruction is 14% at $10^{19}$ eV, determined through complete shower simulation and reconstruction using Auger $\overline{\rm Off}\underline{\rm line}$. The systematic uncertainty from event reconstruction is dominated by the systematic uncertainty of S (1000), with only a few percent coming from the uncertainty of $X^{\rm v}_{\rm max}$ and the zenith angle.

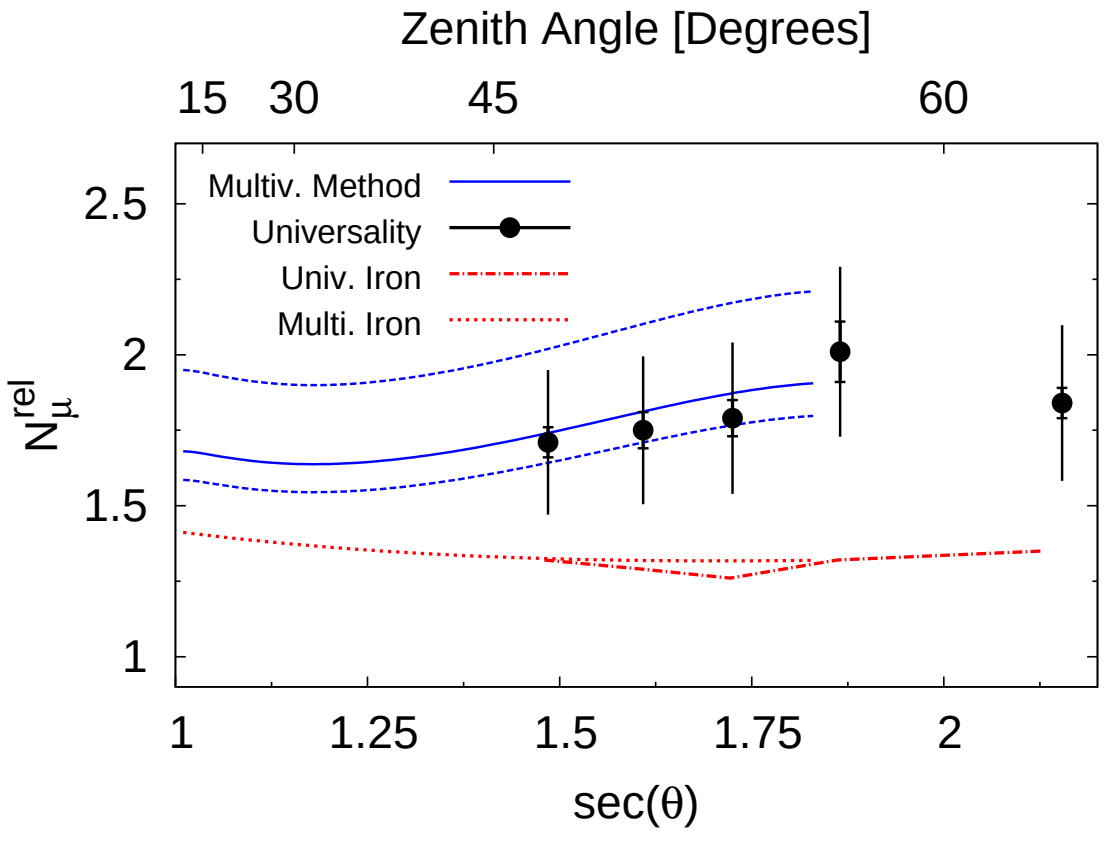

Figure 2: The number of muons estimated at 1000 m in data relative to the predictions of simulations using QGSJET II with proton primaries. The results obtained using the multivariate method are shown as the solid line with systematic uncertainties as the dashed lines. The results obtained using the universality of $S_\mu/S_{\rm em}$ are shown as circles with statistical and systematic uncertainties. The results when the methods are applied on a library of iron-initiated showers are shown as the dot and dash-dot lines.

### 3.3 Application to data

By applying the multivariate and the universality methods to data collected between 1 January 2004 and 30 September 2010, a significant excess of muons is measured compared to the predictions of simulations using QGSJET II; see Fig. 2. The multivariate method was applied to SD events over the energy range $18.6 < \log{(E)} < 19.4$ and $0° - 57°$. The universality method was applied to hybrid events over the energy range $18.8 < \log{(E)} < 19.2$ and $45° - 65°$. For both methods, the excess is estimated here relative to showers simulated with proton primaries.

The multivariate method is used to determine the number of muons, $N_\mu(1000)$. The relative excess is angle-independent, to within 3%, until about 40°, above which it increases. In particular, at $\theta = 38°$ the excess is $\big(1.65^{+0.26}_{-0.10}\big)$ and at $\theta = 55°$ the excess is $\big(1.88^{+0.17}_{-0.06}\big)$. The universality of $S_\mu/S_{\rm em}$ for fixed $X^{\rm v}_{\rm max}$ is used to estimate the total muonic signal. For $45° - 53°$, the relative excess is $(1.76 \pm 0.04({\rm stat.}) \pm 0.29\,({\rm syst.}))$, while for $53° - 65°$ the discrepancy rises to $(1.89 \pm 0.04 \pm 0.28)$.

## 4 Discussion

As demonstrated in the analyses, and shown in Figs. 1 and 2, simulations of air showers using QGSJET II with pro-ton and iron primaries underestimate both the total detector signal at ground level and the number of muons in events collected at the Pierre Auger Observatory. These discrepancies could be caused by an incorrect energy assignment within the 22% systematic uncertainty of the energy scale of the Auger Observatory and/or shortcomings in the simulation of the hadronic and muonic shower components.

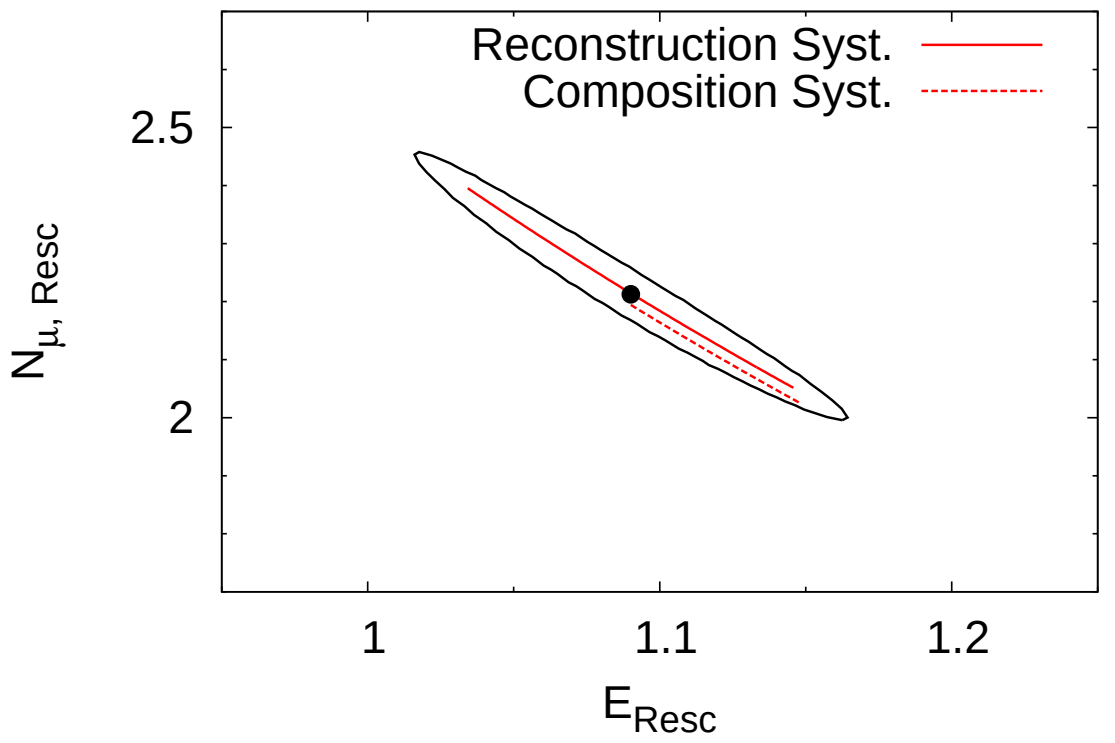

Figure 3: The one-sigma contour of the fit to $N_{\mu,\text{Resc}}$ and $E_{\text{Resc}}$ from the simple matching of the ground signals in simulated and measured hybrid events. The systematic uncertainties from reconstruction and composition are shown as solid and, slightly offset, dashed lines.

To explore these potential sources of discrepancy, a simple modification of the ground signals was implemented in the simulated hybrid events of Sec. 2. The uncertainty in the energy scale motivates the rescaling of the total ground signal by a factor $E_{\text{Resc}}$, and the muon deficit motivates a rescaling of the signal from hadronically produced muons by a factor $N_{\mu,\text{Resc}}$. The rescaled muonic and EM components of $\text{S}\,(1000)$, $S_\mu$ and $S_{\text{EM}}$ – both defined for proton primaries – modify the ground signal

$$\text{S}\,(1000)_{\text{Sim}} = E_{\text{Resc}}^{0.92}\, N_{\mu,\text{Resc}}\, S_{\mu,\text{Sim}} + E_{\text{Resc}}\, S_{\text{EM,Sim}}\ , \quad (4)$$

where the exponent 0.92 is the energy scaling of the muonic signal predicted by simulations. The rescaling factors were applied uniformly to all events. This represents a simplistic modification, and $N_{\mu,\text{Resc}}$ does not reflect any changes in the attenuation and lateral distribution of muons. However, both the attenuation and LDF would change if, for example, the energy spectrum of muons predicted by simulations is not in agreement with the data.

$E_{\text{Resc}}$ and $N_{\mu,\text{Resc}}$ were determined simultaneously by making a maximum-likelihood fit between the modified, simulated $\text{S}\,(1000)$ and the measured $\text{S}\,(1000)$ for the ensemble of hybrid events. The best fit values of $N_{\mu,\text{Resc}}$ and $E_{\text{Resc}}$ are $\left(2.21 \pm 0.23\,(\text{stat.})\ ^{+0.18}_{-0.23}\,(\text{syst.})\right)$ and $\left(1.09 \pm 0.08\ ^{+0.08}_{-0.06}\right)$ respectively; see Fig. 3. The systematic uncertainties arise from uncertainty in the composition and event reconstruction.

The signal rescaling in simulated hybrid events is fundamentally different from the other methods. The observational muon enhancement, $N_\mu^{\text{rel}}$, which includes all muons, cannot be compared directly to $N_{\mu,\text{Resc}}$, which represents an increase of only the hadronically produced muons and their decay products. In addition, the potential increase of $N_\mu^{\text{rel}}$ with zenith angle suggests that a global rescaling of the ground signal from muons is overly simplistic.

In summary, all of the analyses show a significant deficit in the number of muons predicted by simulations using QGSJET II with proton primaries compared to data. This discrepancy cannot be explained by the composition alone, although a heavy composition could reduce the relative excess by up to 40%. The purely-observational estimation of the muonic signal in data, using the signal traces of surface detectors and universal properties of air showers, is compatible with results previously presented and the results obtained from inclined showers [4, 15, 16]. The increased sophistication of the methods gives further weight to the previous conclusions: at the current fluorescence energy scale, the number of muons in data is nearly twice that predicted by simulations of proton-induced showers. The update and application to recent data of the constant intensity cut with universality method and the "smoothing method" are in progress. The possible zenith angle dependence of $N_\mu^{\text{rel}}$ suggests that, in addition to the number, there may also be a discrepancy in the attenuation and lateral distribution of muons between the simulations and data.

An extension of the studies using EPOS 1.99 is in progress.

## References

[1] R. Ulrich, R. Engel, M. Unger, Phys. Rev. D, 2011, **83**: 054026.

[2] T. Pierog, K. Werner, Phys. Rev. Lett., 2008, **101**: 171101.

[3] S. Ostapchenko, Nucl. Phys. Proc. Suppl., 2006, **151**: 143.

[4] A. Castellina, for the Pierre Auger Collaboration, Proc. 31th ICRC, Łódź, Poland, 2009. arXiv:0906.2319v1 [astro-ph].

[5] The Pierre Auger Collaboration, Phys. Rev. Lett., 2008, **101**: 061101.

[6] H.-J. Drescher, G. R. Farrar, Phys. Rev. D, 2003, **61**: 116001.

[7] G. Battistoni et al., AIP Conference Proc., 2007, **896**: 31.

[8] S. Argiro et al., Nucl. Instrum. Meth., 2007, **A580**: 1485.

[9] J. Allen et al., J. Phys. Conf. Ser., 2008, **119**: 032002.

[10] I. Nabney: 2002, Netlab: Algorithms for Pattern Recognition, Springer.

[11] A. Yushkov et al., Phys. Rev. D, 2010, **81**: 123004.

[12] D. D'Urso et al., paper 0694, these proceedings.

[13] A. Yushkov et al., paper 0687, these proceedings.

[14] D. Heck et al., FZKA 6019, 1998, Forschungzentrum Karlsruhe.

[15] R. Engel, for the Pierre Auger Collaboration, Proc. 30th ICRC, Merida, Mexico, 2007, **4**: 385.

[16] G. Rodriguez, for the Pierre Auger Collaboration, paper 0718, these proceedings.

# A new method for determining the primary energy from the calorimetric energy of showers observed in hybrid mode on a shower-by-shower basis

ANALISA G. MARIAZZI[1] FOR THE PIERRE AUGER COLLABORATION[2]
[1]*IFLP, Universidad Nacional de La Plata and CONICET, La Plata, Argentina*
[2]*Observatorio Pierre Auger, Av. San Martin Norte 304, 5613 Malargüe, Argentina*
*(Full author list: http://www.auger.org/archive/authors_2011_05.html)*
*auger_spokespersons@fnal.gov*

**Abstract:** The energy deposit integral can be used for determining the calorimetric energy of air showers observed with fluorescence telescopes. The invisible fraction of the primary energy, averaged over many showers, is typically estimated from Monte Carlo simulations and later added for the reconstruction of the total primary energy. In this contribution we derive a simple parameterization of the invisible energy correction that can be applied to individual events measured with both the fluorescence and the surface detectors. The obtained parameterization is robust with respect to a change of the high energy hadronic interaction model employed in the simulation and has only a very small primary mass dependence, reducing the associated systematic uncertainties of energy reconstruction.

**Keywords:** Ultra High Energy Extensive Air Showers, Missing Energy, Muons, Hadronic interactions

## 1 Introduction

When an ultra high energy cosmic ray interacts in the atmosphere a cascade of particles is generated. In the cascade, an important fraction of the energy is deposited in the atmosphere as ionization of the air molecules and atoms, and the remaining fraction is carried away by neutrinos and high energy muons that hit the ground.

A fraction of the total deposited energy is re-emitted during the de-excitation of the ionized molecules as fluorescence light that can be detected by fluorescence telescopes. The telescopes use the atmosphere as a calorimeter, making a direct measurement of the longitudinal shower development. The energy deposit integral can be used to determine the calorimetric energy ($E_{Cal}$) of air showers observed with fluorescence telescopes.

The fraction of energy carried away by neutrinos and high energy muons is *a priori* unknown, and corrections for this so-called missing energy ($E_{Missing}$) must be properly applied to the measured $E_{Cal}$ to find the primary energy ($E_{Primary}$). Generally, the missing energy correction is parameterized as a function of $E_{Cal}$ ($E_{Missing}(E_{Cal})$), which is estimated from Monte Carlo simulations averaging over many showers. The missing energy is about 10% of the primary energy depending on the high energy hadronic interaction model and on the primary mass, as shown in Figure 1. Since the primary mass cannot be determined on an event by event basis, an average mass composition must be assumed. This introduces a systematic uncertainty in the determination of the primary energy and possibly a bias, if the actual mass composition is different from the assumed average.

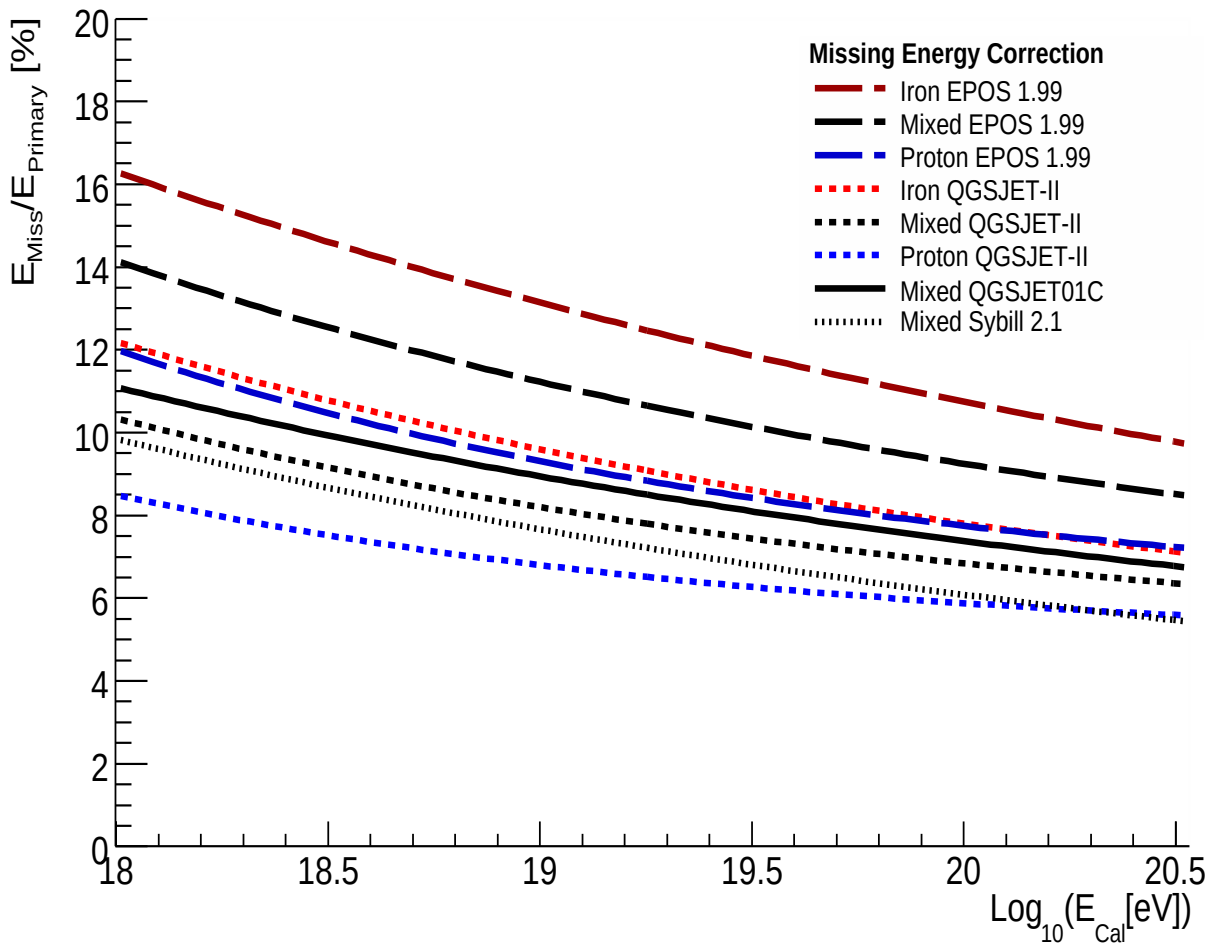

Figure 1: $E_{Missing}(E_{Cal})$ correction for fluorescence detectors.

The model dependence of the missing energy estimation as a function of $E_{Cal}$ is a direct consequence of using a parameter that is not actually related to the missing energy, but to the electromagnetic energy. The lack of knowledge of the correct interaction model at high energies also introduces a systematic uncertainty and possibly a bias, that are ultimately not known. For example, a mis-reconstruction of the missing energy due to appearance of new physics in the hadronic interaction models could explain features of the cosmic ray energy spectrum like the knee without making any use of changes in the primary spectrum slope[1]. The event by event deviations with respect to a reference average value of the missing energy for QGSJET01 mixed mass composition($E^{QGSJET01}_{Missing}(E_{Cal})$ ) can be seen in Figure 2 . Looking at the spread caused by the misreconstruction of the missing energy, it is then desirable to have a missing energy parameterization as a function of shower observables that are less model dependent, and give a better estimation of the true missing energy on a shower-by-shower basis.

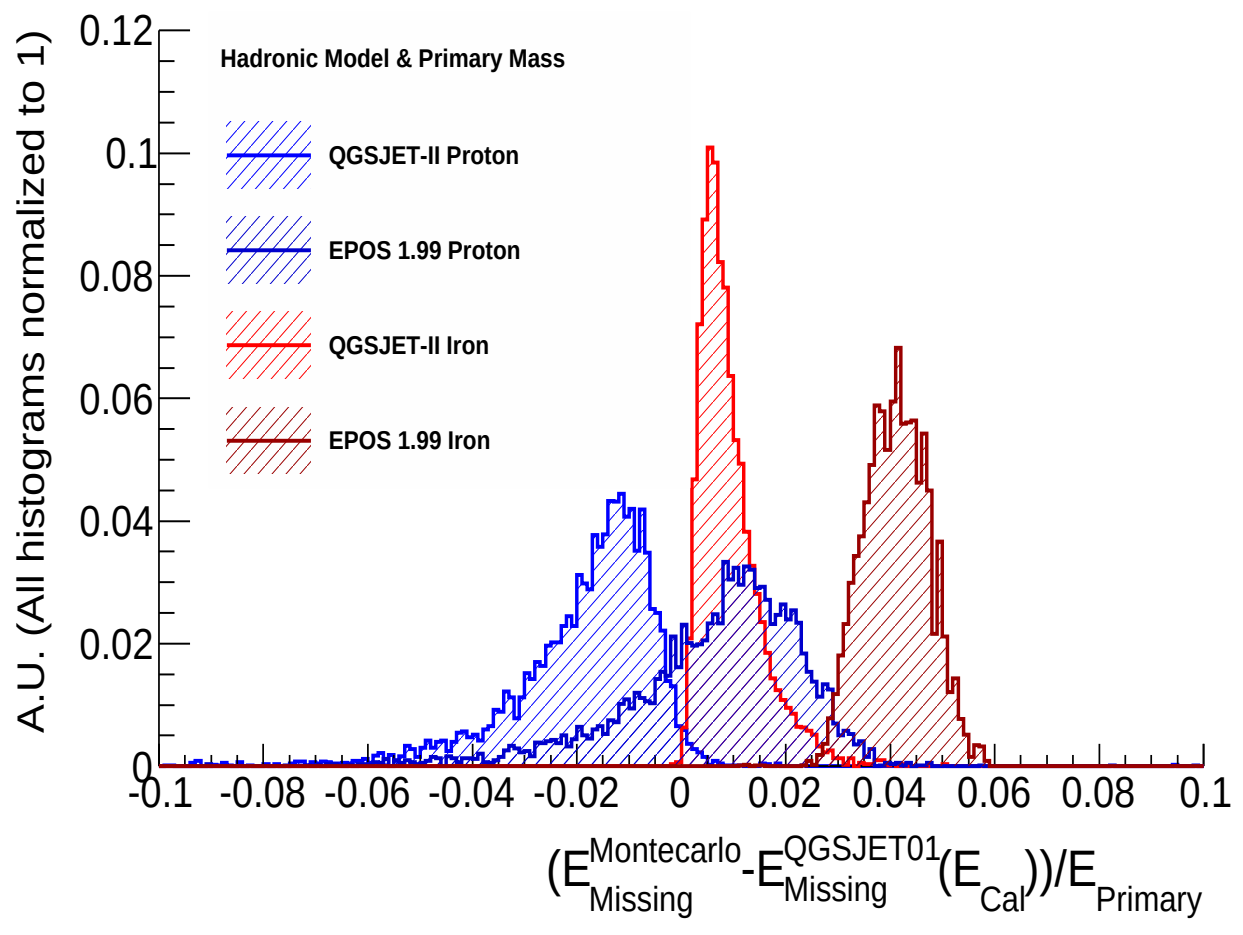

Figure 2: Difference between $E^{QGSJET01}_{Missing}(E_{Cal})$ and the missing energy from showers simulated with EPOS and QGSJETII, in units of the primary energy.

Extensive air showers created by ultra-high energy cosmic ray are measured with two complementary techniques at the Pierre Auger Observatory. The longitudinal shower development is recorded with the Fluorescence Detector (FD), while the muonic and electromagnetic components can be measured at ground by the Surface Detector (SD). The lateral distribution of the shower particles at ground is sampled with an array of more than 1600 water-Cherenkov detectors while the fluorescence light emission along the shower trajectory through the atmosphere is observed with a set of 24 telescopes [2].

In this article, a new approach for the determination of the missing energy in extensive air showers is presented. This approach takes advantage of the hybrid nature of the Pierre Auger Observatory, using the signal at 1000 m from the shower core ($S(1000)$) and the atmospheric slant depth of the shower maximum ($X_{max}$) to provide a robust estimation of the missing energy, reducing the systematic uncertainties that this correction introduces in the determination of the primary energy.

## 2 A Toy Model for the Missing Energy

In the Heitler model extended to hadronic cascades by Matthews [3], the primary energy is distributed between electromagnetic particles and muons.

$$E_0 = \xi^e_c N_{max} + \xi^\pi_c N_\mu \tag{1}$$

One can identify the second term directly as the missing energy:

$$E_{Missing} = \xi^\pi_c\, N_\mu \tag{2}$$

where $\xi^e_c$ is the critical energy for the electromagnetic particles and $\xi^\pi_c$ is the pion critical energy. Although the number of muons generated in the shower depends on the high energy hadronic interaction model, the pion critical energy is a well established quantity that depends primarily on the medium density where the first interactions take place, making this relationship robust to changes in the hadronic interaction model.

Nyklicek et al. [4] have shown using vertical showers that the model dependence is reduced if $E_{Missing}$ is estimated using its correlation with the total number of muons at ground above 1 GeV. In their work there is a linear relation to the number of muons (Eq. 2) and the proportionality factor obtained from Monte Carlo simulations is of the order of a critical energy $\xi^\pi_c \approx 10$ GeV which is in agreement with the predictions made using the extended toy model of Matthews [3].

An observable related to the muon content of the shower would be more suitable for the determination of the missing energy correction. However, the number of muons is not directly measured in the Pierre Auger Observatory. One of the simplest observables related to the muon content of the shower is $S(1000)$.

Based on universality studies [5, 6], the relationship between $S(1000)$ and the muon content should be universal when expressed as a function of the stage of development of the cascade at ground level measured by $DX = X_{ground} - X_{max}$ (distance measured in atmospheric depth from the ground to the point of maximum development of the shower). For a fixed $DX$, a change in the primary mass or the hadronic model that modifies the muon content of the shower (an thus, the missing energy) will change $S(1000)$ accordingly. This makes the combination of these parameters more robust for the determination of the missing energy, and less dependent of the details of the hadronic interactions or the primary mass composition. Even if the Heitler model is an oversimplification, it provides great insight on the phenomenology of shower cascades. The total number of muons follows a power law with the primary energy

$$N_\mu = \left(\frac{E_0}{\xi^\pi_c}\right)^\beta . \tag{3}$$

The primary energy $E_0$ is also a power law of $S(1000)$ for a fixed angle ($S_{38^\circ}$), or for a fixed stage of shower devel-

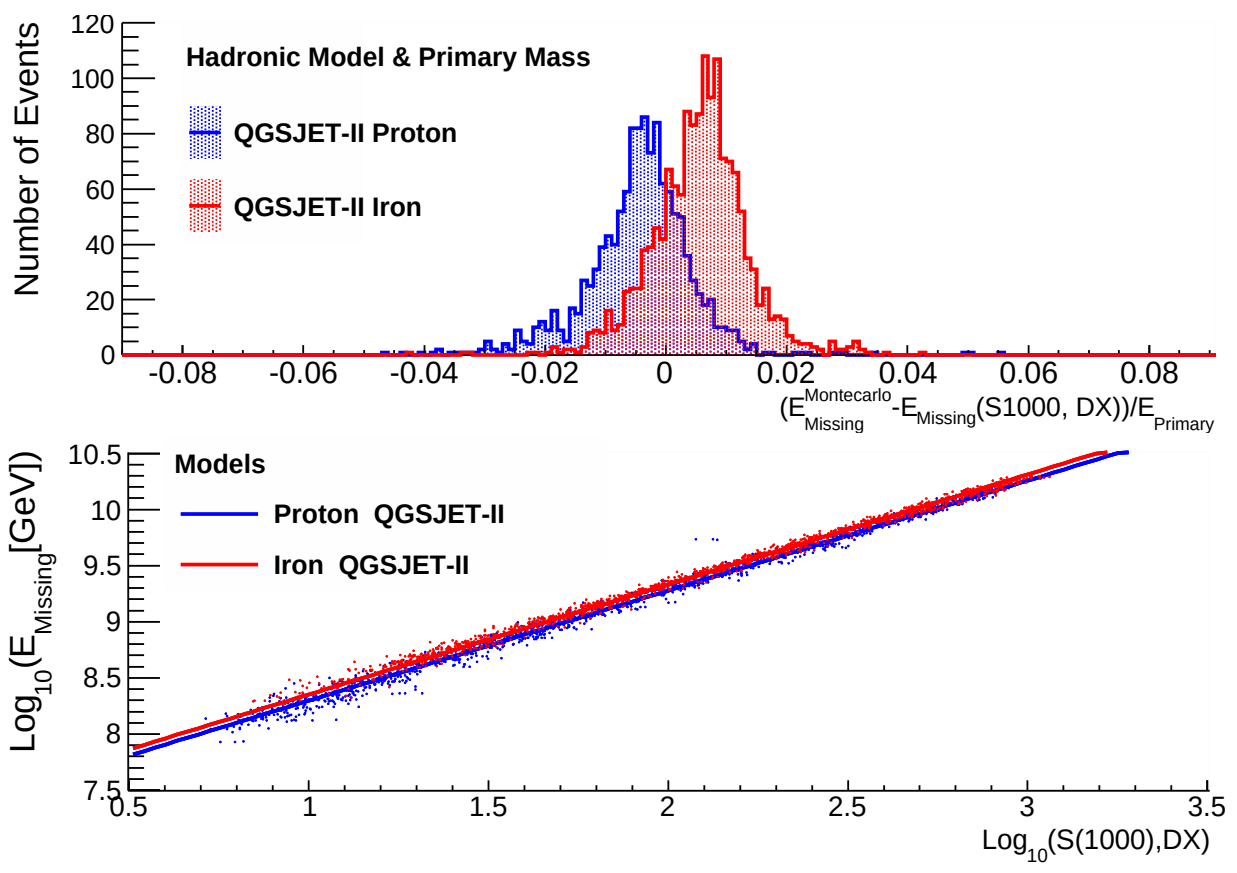

Figure 3: Fit of $\log(E_{Missing}[GeV])$ vs. $\log(S(1000)[VEM])$(bottom) and its residues(top) for fixed $DX$ bin.

opment using universality in $DX$

$$N_\mu = \left(\frac{\alpha(DX)S(1000)^\gamma}{\xi_c^\pi}\right)^\beta \quad (4)$$

where the function $\alpha(DX)$ takes into account the attenuation with $DX$. Based on this toy model, one can estimate the missing energy using $S(1000)$ and $DX$

$$\begin{aligned} \log(E_{Missing}) &= A(DX) + B\,\log(S(1000)) \quad (5)\\ A(DX) &= \log(\xi_c^\pi) + \beta\log\left(\frac{\alpha(DX)}{\xi_c^\pi}\right)\\ B &= \beta\gamma. \end{aligned}$$

$A(DX)$ and $B$ will have to be determined with fits to Monte Carlo simulations. As we will see in section 3, the $B$ parameter can be set to a fixed value close to unity. This is easy to understand if we consider that in this simple model $\beta$ depends on the inelasticity and multiplicity of pion interactions and is usually within 10% of 0.9 [3] and $\gamma$ is in the 1.06 - 1.09 range [7]. Once the values of A and B are known, we will be able to estimate the missing energy of any event where $S(1000)$ and $X_{max}$ are measured. We will call this new parameterization of the missing energy $E_{Missing}(S(1000), DX)$.

## 3 Results and conclusions

Showers simulated with CORSIKA[8] were subsequently used as input in the detector simulation code, and reconstructed using the official Offline reconstruction framework of the Pierre Auger Observatory [9]. The generated data sample contains approximately $4\times10^4$ showers simulated using the hadronic interaction model QGSJETII(03)[10]. This library consists of proton and iron initiated showers following a power law primary energy spectrum ($E^{-1}$) in the energy range $\log(E/eV)$ = 18.5 - 20.0 and uniformly distributed in $\cos^2\theta$ in zenith angle range $\theta = 0-65°$. The EPOS 1.99 [11] generated data sample contains also approximately $4\times10^4$ showers but discrete values of energy and zenith angle.

The $X_{max}$ value for Monte Carlo simulations was taken from the Gaisser Hillas fit of the longitudinal energy deposit profile and the missing energy of the simulated event was calculated following [12]. Since the simulations in the library are not hybrid, the FD reconstruction accuracy was factored in by introducing a 20% Gaussian smearing of the Monte Carlo calorimetric energy, a 2° Gaussian smearing of the primary zenith angle and a 25 $\mathrm{g\,cm^{-2}}$ Gaussian smearing of $X_{max}$. These values are rather conservative for hybrid events and the results presented in this work are insensitive to the value of these parameters, as long as they are kept within a reasonable range.

The shower library generated with the QGSJETII hadronic interaction model was used to parameterize the missing energy as a function of $S(1000)$ and $DX$. The surface detector events had to satisfy quality cuts for good $S(1000)$ reconstruction[13]. The showers were divided in 13 equidistant bins of $DX$, ranging from 175 to 1100 $\mathrm{g\,cm^{-2}}$. For each bin of $DX$, the missing energy is fitted using equation (5). A representative example of these fits and the corresponding residues are shown in figure 3. The variation of the parameter $A$ with $DX$ was then parameterized with a third degree polynomial

$$\begin{aligned} A(DX) = 7.347 - 3.41\,10^{-4}DX + 1.58\,10^{-6}DX^2\\ -7.88\,10^{-10}DX^3 \quad (6) \end{aligned}$$

the parameter $B$ was fixed to 0.98 and the $A$ parameter dependence with $DX$ for QGSJETII showers. The difference between $E_{Missing}(S(1000), DX)$ and the actual missing energy of the QGSJETII showers as a function of $E_{Cal}$ is presented in Figure 4 (left). Filled circles represent the values obtained using $E_{Missing}(S(1000), DX)$ and empty circles represent Monte Carlo true values, which are slightly shifted to the left to aid clarity. There is a good agreement with a small bias of less than 1 % of the primary energy depending on the mass composition and its value decreases with primary energy. The set of EPOS simulations was used to test how $E_{Missing}(S(1000), DX)$ performed with a change in the hadronic model. EPOS is significantly different from QGSJETII and is known to generate more muons than other models, and consequently more missing energy. The difference between $E_{Missing}(S(1000), DX)$ and the actual missing energy of the EPOS showers is presented in Figure 4(right). It is important to emphasise that we are using a parameterization obtained from QGSJETII showers to describe the missing energy given by a significantly different hadronic model like EPOS, without introducing important biases or loosing too much accuracy. As we mentioned in the introduction, this is possible because we are estimating the missing energy using observables closely related to the muonic component of the shower at a given shower development

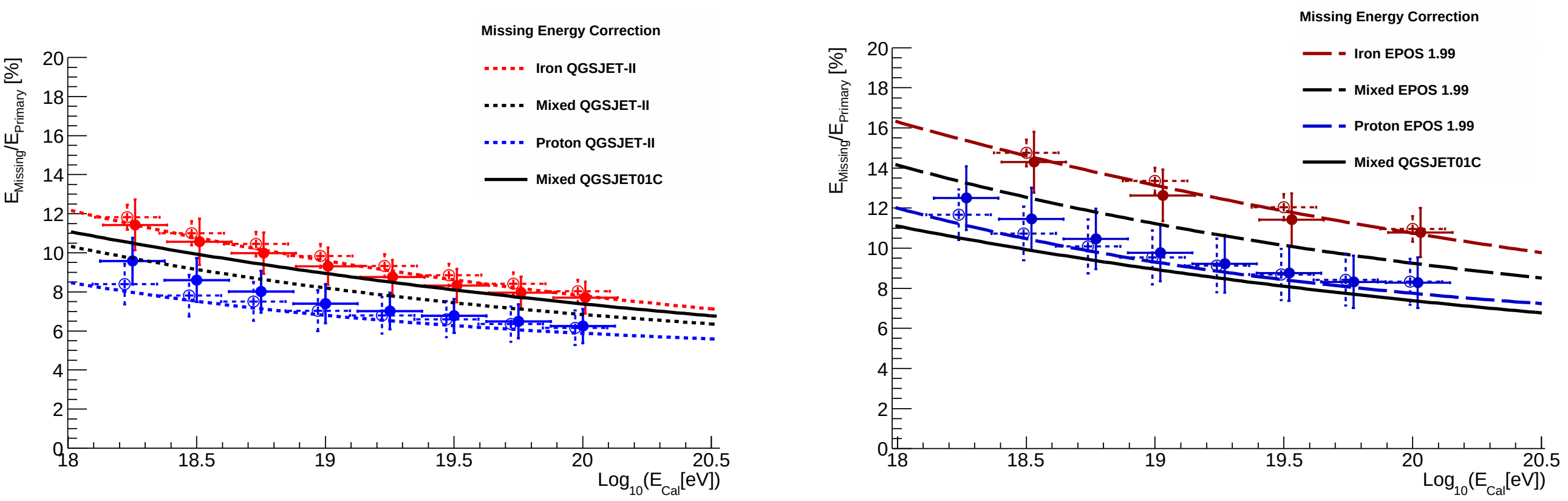

Figure 4: Average missing energy $E_{Missing}(S1000, DX)$ (filled circles) as a function of the calorimetric energy for QGSJETII (left) and for EPOS (right) using the $E_{Missing}(S1000, DX)$ from QGSJETII. Monte Carlo true values (empty circles) are slightly shifted to the left to aid clarity. Lines represent $E_{Missing}(E_{Cal})$ parameterizations.

stage, that are in turn tightly related to the origin of the missing energy.

To illustrate this point, in figures 2 and 5 we show the difference between the missing energy parameterizations and the simulation true values for each of the considered parameterizations. It can be seen in figure 5 how $E_{Missing}(S(1000), DX)$ gives a better estimation of the missing energy than $E^{QGSJET01}_{Missing}(E_{Cal})$, even if the hadronic model or the primary mass is changed. Using the presented missing energy estimator $E_{Missing}(S(1000), DX)$, the hadronic interaction model bias is removed while, at the same time, the bias due to the mass composition is reduced by a factor of two with respect to the previous $E^{QGSJET01}_{Missing}(E_{Cal})$ parameterization.

$E_{Missing}(S(1000), DX)$ enables us to estimate the missing energy of almost any event with a good reconstruction of $S(1000)$ and $X_{max}$, without making assumptions on the primary mass or the hadronic model. Future work will also include tests with other hadronic interaction models to strengthen the hypothesis of hadronic model independence, and an extension applicable to the reconstruction of very inclined showers.

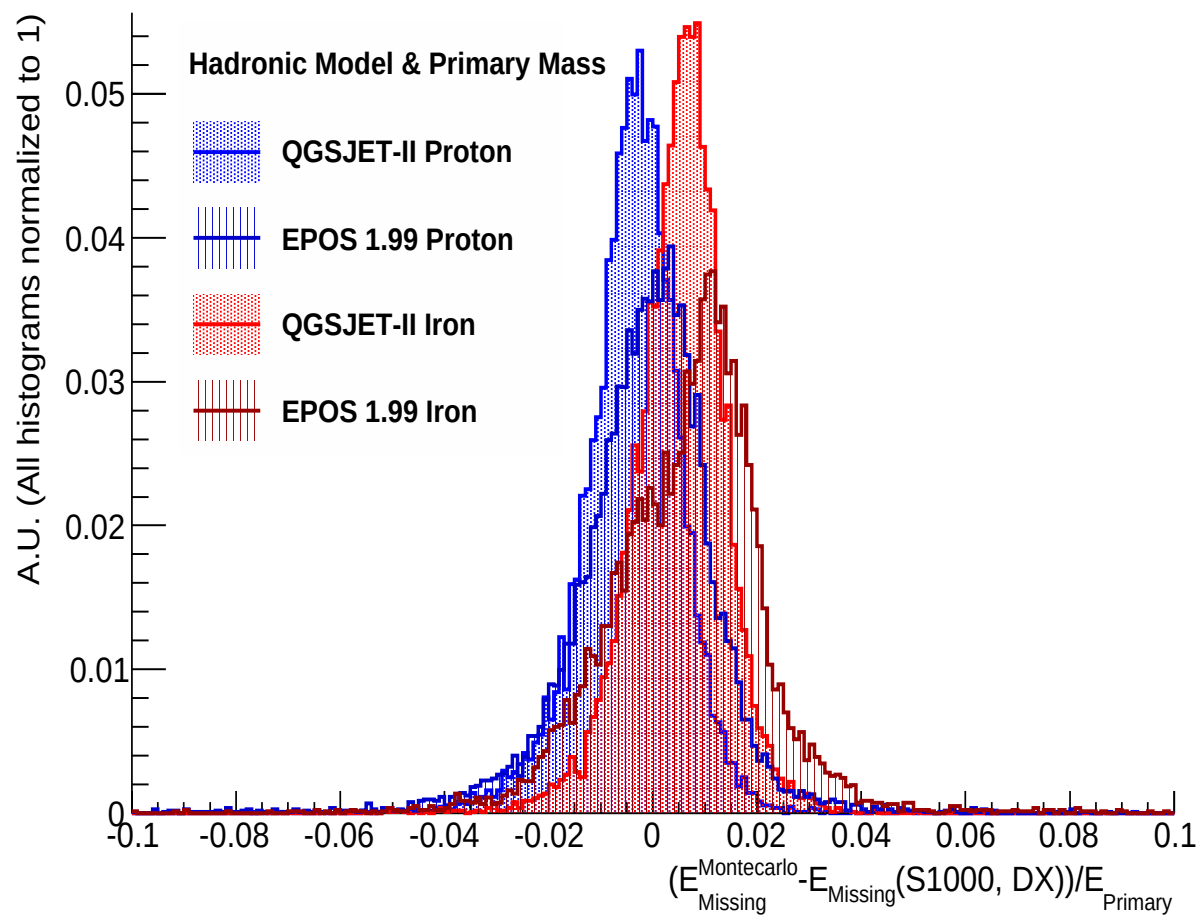

Figure 5: Difference between $E_{Missing}(S(1000), DX)$ and the missing energy from showers simulated with EPOS and QGSJETII, in units of the primary energy.

Hybrid events that trigger the surface detector array and the fluorescence telescopes separately are ideally suited to estimate the missing energy. The application of this method to determine the missing energy from a set of such hybrid events and a detailed study of the impact of this new missing energy correction on the surface detector calibration are in progress.

## References

[1] A. Petrukhin, Nucl. Phys. B (Proc. Suppl.), 2006, **151**: 57.

[2] The Pierre Auger Collaboration, Nucl. Instrum. Meth. A, 2004, **523**: 50.

[3] J. Matthews, Astropart. Phys., 2005, **22**: 387.

[4] M. Nyclicek et al., Proc. 31st ICRC, Łódź, Poland, 2009.

[5] A. Yushkov et al., Phys. Rev. D, 2010, **81**: 123004.

[6] F. Schmidt et al., Astropart. Phys., 2008, **29**(Issue 6): 355.

[7] The Pierre Auger Collaboration, Phys. Rev. Lett., 2008, **101**:061101.

[8] D. Heck et al., FZKA 6019, Forschungszentrum Karlsruhe(1998),http://www-ik.fzk.de/corsika.

[9] S. Argiro et al, Nucl. Instrum. Meth. A, 2007, **580**: 1485.

[10] S. Ostapchenko, Phys. Rev. D, 2006, **74**: 014026.

[11] K. Werner et al., Nucl. Phys. Proc. Suppl., 2008, **175**: 81.

[12] H. Barbosa et al., Astropart. Phys., 2004, **22**: 159.

[13] The Pierre Auger Collaboration, Nucl. Instrum. Meth. A, 2010, **613**: 29.

# Acknowledgments

The successful installation, commissioning and operation of the Pierre Auger Observatory would not have been possible without the strong commitment and effort from the technical and administrative staff in Malargüe.

We are very grateful to the following agencies and organizations for financial support:

Comisión Nacional de Energía Atómica, Fundación Antorchas, Gobierno De La Provincia de Mendoza, Municipalidad de Malargüe, NDM Holdings and Valle Las Leñas, in gratitude for their continuing cooperation over land access, Argentina; the Australian Research Council; Conselho Nacional de Desenvolvimento Científico e Tecnológico (CNPq), Financiadora de Estudos e Projetos (FINEP), Fundação de Amparo à Pesquisa do Estado de Rio de Janeiro (FAPERJ), Fundação de Amparo à Pesquisa do Estado de São Paulo (FAPESP), Ministério de Ciência e Tecnologia (MCT), Brazil; AVCR AV0Z10100502 and AV0Z10100522, GAAV KJB100100904, MSMT-CR LA08016, LC527, 1M06002, and MSM0021620859, Czech Republic; Centre de Calcul IN2P3/CNRS, Centre National de la Recherche Scientifique (CNRS), Conseil Régional Ile-de-France, Département Physique Nucléaire et Corpusculaire (PNC-IN2P3/CNRS), Département Sciences de l'Univers (SDU-INSU/CNRS), France; Bundesministerium für Bildung und Forschung (BMBF), Deutsche Forschungsgemeinschaft (DFG), Finanzministerium Baden-Württemberg, Helmholtz-Gemeinschaft Deutscher Forschungszentren (HGF), Ministerium für Wissenschaft und Forschung, Nordrhein-Westfalen, Ministerium für Wissenschaft, Forschung und Kunst, Baden-Württemberg, Germany; Istituto Nazionale di Fisica Nucleare (INFN), Ministero dell'Istruzione, dell'Università e della Ricerca (MIUR), Italy; Consejo Nacional de Ciencia y Tecnología (CONACYT), Mexico; Ministerie van Onderwijs, Cultuur en Wetenschap, Nederlandse Organisatie voor Wetenschappelijk Onderzoek (NWO), Stichting voor Fundamenteel Onderzoek der Materie (FOM), Netherlands; Ministry of Science and Higher Education, Grant Nos. 1 P03 D 014 30, N202 090 31/0623, and PAP/218/2006, Poland; Fundação para a Ciência e a Tecnologia, Portugal; Ministry for Higher Education, Science, and Technology, Slovenian Research Agency, Slovenia; Comunidad de Madrid, Consejería de Educación de la Comunidad de Castilla La Mancha, FEDER funds, Ministerio de Ciencia e Innovación and Consolider-Ingenio 2010 (CPAN), Xunta de Galicia, Spain; Science and Technology Facilities Council, United Kingdom; Department of Energy, Contract Nos. DE-AC02-07CH11359, DE-FR02-04ER41300, National Science Foundation, Grant No. 0450696, The Grainger Foundation USA; ALFA-EC / HELEN, European Union 6th Framework Program, Grant No. MEIF-CT-2005-025057, European Union 7th Framework Program, Grant No. PIEF-GA-2008-220240, and UNESCO.